\documentclass[lettersize,journal]{IEEEtran}
\IEEEoverridecommandlockouts
\usepackage{hyperref}
\usepackage{url}
\usepackage{tikz}
\usepackage{caption}
\usetikzlibrary{positioning, shapes.geometric, arrows.meta, fit}
\usetikzlibrary{shapes.geometric, arrows.meta}
\tikzstyle{blockish} = [rectangle, minimum width=3cm, minimum height=1cm, text centered, draw=black, fill=blue!30]
\tikzstyle{startstop} = [rectangle, rounded corners, minimum width=3cm, minimum height=1cm,text centered, draw=black, fill=red!30]
\tikzstyle{process} = [rectangle, minimum width=3cm, minimum height=1cm, text centered, draw=black, fill=blue!30]
\tikzstyle{decision} = [diamond, minimum width=3cm, minimum height=1cm, text centered, draw=black, fill=green!30]

\tikzstyle{block} = [rectangle, rounded corners, minimum width=2.5cm, minimum height=1cm, text centered, draw=black, fill=green!30]
\tikzstyle{block2} = [rectangle, rounded corners, minimum width=2.5cm, minimum height=1cm, text centered, draw=black, fill=blue!30]
\tikzstyle{block3} = [rectangle, rounded corners, minimum width=2.5cm, minimum height=1cm, text centered, draw=black, fill=red!30]
\tikzstyle{block_start_end} = [rectangle, rounded corners, minimum width=2.5cm, minimum height=1cm, text centered, draw=black, fill=gray!30]
\tikzstyle{arrow} = [thick,->,>=stealth]

\usepackage{cite}
\usepackage{balance}
\usepackage{comment}
\usepackage{amsmath,amssymb,amsfonts}
\usepackage{algorithmic}
\usepackage{tcolorbox}
\usepackage{graphicx}
\usepackage{multirow}
\usepackage{amsmath}
\usepackage{booktabs}
\usepackage{textcomp}
\usepackage{xcolor}
\def\BibTeX{{\rm B\kern-.05em{\sc i\kern-.025em b}\kern-.08em
    T\kern-.1667em\lower.7ex\hbox{E}\kern-.125emX}}
    
\begin{document}

\title{A Robot-Led Intervention for Emotion Regulation: From Expression to Reappraisal}

\author{\large Guy Laban$^{1,*}$, \and Julie Wang$^{2}$, \and Hatice Gunes$^1$\\
 {\normalsize
    $^1$ Department of Computer Science and Technology, University of Cambridge, Cambridge, United Kingdom\\
    $^2$ Department of Psychology, University of Cambridge, Cambridge, United Kingdom.}
\thanks{* Corresponding Author 
    {\tt guy.laban@cl.cam.ac.uk}}%
\thanks{G. Laban and H. Gunes are supported by the EPSRC project ARoEQ under grant ref. EP/R030782/1. J. Wang undertook this work while she was a visiting undergraduate student at the AFAR Lab, Department of CST, University of Cambridge.} \thanks{\textbf{Open Access:} For open access purposes, the authors have applied a Creative Commons Attribution (CC BY) licence to any Author Accepted Manuscript version arising. \textbf{Data access:} Raw data related to this publication cannot be openly released due to anonymity and privacy issues.}
}

\maketitle

\begin{abstract}
Emotion regulation is a crucial skill for managing emotions in everyday life, yet finding a constructive and accessible method to support these processes remains challenging due to their cognitive demands. In this study, we explore how regular interactions with a social robot, conducted in a structured yet familiar environment within university halls and departments, can provide effective support for emotion regulation through cognitive reappraisal. Twenty one students participated in a five-session study at a university hall or department, where the robot (QTrobot, LuxAI), powered by a large language model (GPT 3.5) facilitated structured conversations, encouraging the students to reinterpret emotionally charged situations that they shared with the robot. Quantitative and qualitative results indicate significant improvements in emotion self-regulation, with participants reporting better understanding and control of their emotions. The intervention led to significant changes in constructive emotion regulation tendencies and positive effects on mood and sentiment after each session. The findings also demonstrate that repeated interactions with the robot encouraged greater emotional expressiveness, including longer speech disclosures, increased use of affective language, and heightened facial arousal. Notably, expressiveness followed structured patterns aligned with the reappraisal process, with expression peaking during key reappraisal moments, particularly when participants were prompted to reinterpret negative experiences. The qualitative feedback further highlighted how the robot fostered introspection and provided a supportive space for discussing emotions, enabling participants to confront long-avoided emotional challenges. These findings demonstrate the potential of robots to effectively assist in emotion regulation in familiar environments, offering both emotional support and cognitive guidance.
\end{abstract}

\begin{IEEEkeywords}
Human--Robot Interaction, Emotion Regulation, Cognitive Reappraisal, Well-being, Longitudinal, Large Language Models
\end{IEEEkeywords}
\section{Introduction}
University students face significant emotional challenges as they balance academic pressures, personal growth, and social expectations \cite{Park2020UnderstandingStudy.}. These challenges can result in heightened stress, anxiety, and emotional turmoil, making it difficult for students to manage their emotions effectively \cite{CordovaOlivera2023AcademicStudents}. Without practicing \textit{emotion regulation}, these emotional experiences can negatively impact their well-being, academic performance \cite{Andres2017EmotionRelationships}, and overall mental health \cite{Sheppes2015}. Emotion regulation, a set of 
processes, strategies and techniques involving the monitoring, assessment, and modification of one’s state, behaviour, or cognition in a given situation, serves as a vital means of coping with emotional distress and its associated comorbidities \cite{Gross1998TheReview}. 
These strategies can be either constructive or maladaptive \cite{Garnefski2001NegativeProblems,Gross1993EmotionalBehaviorb,Fearey2021EmotionBehaviours,Carlucci2018Co-ruminationHurt,Fledderus2010}. Constructive strategies require more cognitive effort, making them harder to consistently apply \cite{Gross2014EmotionPerspective,Sheppes2011Emotion-RegulationChoice}. Hence, introducing individuals to constructive emotion regulation skills at a preventive stage is crucial, as it is suggested that when experiencing poor mental health, people tend to adopt maladaptive strategies that perpetuate negative emotional interpretations \cite{Millgram2015SadDepression}. Traditional methods for training and facilitating emotion regulation, such as therapy or peer support, can be effective but are often limited by availability, cost, and accessibility \cite{Boemo2022RelationsAssessments}, especially in high-pressure academic environments \cite{Andres2017EmotionRelationships}. This highlights the need for scalable approaches to support emotion regulation in naturalistic settings. 

Social robots, 
machines that are aimed at interacting  socially with humans 
\cite{breazeal_designing_2004}, offer a potentially scalable and consistent alternative, providing regular, personalized emotional support without constraints of 
time and availability. 
The non-judgmental and objective nature of robots can create a safe socio-emotional space for individuals to express their feelings openly \cite{Laban2024SharingFeel}. Due to social robots' ability to establish rapport \cite{Henschel2021,RefWorks:446}, prior research has shown that robots can engage users in meaningful social interactions \cite{Laban2024SharingFeel} for supporting well being \cite{RefWorks:404,Laban2024SocialWell-Being}. By integrating into students' everyday routines, robots could help bridge gaps in access to emotional regulation tools, making support more readily available and continuous. It 
can reduce the cognitive and emotional barriers that often prevent young individuals from seeking help \cite{Gulliver2010PerceivedReview} and encourage them to discuss their problems openly.
Here, we are specifically interested in the robot's ability to deliver interventions supporting students' \textit{cognitive reappraisal}, a constrictive emotion regulation strategy aimed at changing the way one interprets emotionally charged situations to reduce their negative impact \cite{Gross1998Antecedent-Physiology}, with direct and long-term impact on well-being 
\cite{Webb2012DealingRegulation}. However, it requires substantial cognitive effort to consistently apply, particularly in stressful situations \cite{Sheppes2009ReappraisalEffort,Sheppes2011Emotion-RegulationChoice}. By mediating this process, social robots can provide external guidance, helping students engage in reappraisal more effectively. The robot’s conversational support could ease the cognitive demands by prompting reflection, offering new perspectives, while providing space for talking about their emotions \cite{Laban2024SharingFeel}
, making it easier for students to integrate this crucial skill into their everyday lives. 

In this paper, we present the \textit{first study} exploring how regular interactions with a conversational social robot in familiar environments can support emotional understanding and improve emotion regulation and expression through cognitive reappraisal. Twenty-one university students participated in a multi-session study including five sessions talking with a robot, designed to facilitate cognitive reappraisal intervention. 
The sessions took place in natural settings familiar to the participants, such as their university campus, 
to maintain ecological validity. Each session, scheduled at the participant's convenience via an app, focused on key aspects of well-being, and the robot engaged participants in structured conversations aimed at prompting 
cognitive reappraisal. The robot utilised a large language model (LLM) to process participants' responses and provide supportive, reflective feedback. 
We collected quantitative data using standardized questionnaires and qualitative data from in-person semi-structured interviews, employing a mixed-methods approach to gain a comprehensive understanding of how robot-mediated cognitive reappraisal influences students' emotional understanding, regulation strategies, and overall well-being over time. In addition to self-report measures and qualitative interviews, we analysed the interaction logs and videos to capture verbal expression (e.g., what participants said and how they said it) and facial expression cues (participants’ nonverbal emotional displays) during the sessions. This multi-modal data provided objective indicators of participants’ emotional engagement and expression throughout the intervention.

\section{Related works}
\label{related}

Research on social robots for emotion regulation is still in its early stages, but recent studies provide promising insights into their potential to enhance emotional well-being in various settings. In one study, long-term interactions with a social robot showed that it can foster increased self-disclosure over time via affect labeling (
see \cite{Torre2018PuttingRegulation}), supporting emotional well-being. Participants disclosed more to the robot over time, while  perceiving themselves as being more open \cite{Laban2024BuildingTime}. A replication of this study deployed a social robot as an intervention for informal caregivers experiencing emotional distress. It found that caregivers benefited from self-disclosure to the robot, becoming more accepting of their caregiving role while reporting improved mood and reduced feelings of loneliness \cite{laban_ced_2023}. In a different study, a robot guided participants through Loving-Kindness and Walking Meditation, which were found to significantly increase openness and positive emotions 
\cite{Huang2023Loving-kindnessCreativity}. Another study examined teleoperated social robots used for mindfulness, 
demonstrating their positive impact on relaxation 
over time \cite{Bodala2021TeleoperatedStudy}. 
One study also deployed robotic coaches for facilitating group mindfulness practices in public settings, showing that despite mismatched expectations regarding the robots' capabilities, they were able to promote well-being, social engagement, and adherence to the practices, illustrating their potential to support long-term emotional health even in less-controlled environments \cite{Axelsson2023RoboticCafe}. Other studies have reported on robots' ability to facilitate emotion regulation through means other than conversational interactions (e.g., \cite{Matheus2022ABreathing}). In parallel with these robotic interventions, due to recent advances in artificial intelligence, previous studies have evaluated LLM capacity for emotion regulation. For example, Zhan et al. \cite{Zhan2024LargeGuided} demonstrated that by employing expert-informed reappraisal frameworks, LLMs can generate targeted cognitive reappraisals that outperform human responses in empathy and effectiveness. Similarly, Li et al. \cite{Li2024SkillScenarios} found that GPT-4 surpasses human re-appraisers in dimensions such as effectiveness, empathy, and novelty—with the performance gap driven primarily by intrinsic skill rather than effort.


Young adults and students who face unique emotional challenges could benefit from social robots designed specifically to meet their emotional needs \cite{Sharp2018AIntervention}. A study on intelligent agents supporting students' mental well-being found that university students experience significant stress and anxiety in social situations such as public speaking and group discussions. Students preferred animal-like robots and conversational agents to manage their anxiety, highlighting the potential of these technologies to address mental health challenges \cite{Rasouli2024UniversitySituations}. Non-conversational socially assistive robot (Purrble) was introduced to highly anxious university students as in situ emotional support tools. It demonstrated positive effects, reducing anxiety symptoms and improving students' emotion regulation self-efficacy, making pet-like robots a viable option for managing anxiety in educational contexts \cite{Williams2023FeasibilityTrial}. However, due to Purrble’s pet-like embodiment, the robot's role was primarily focused on providing companionship rather than offering guidance for reappraising emotions. 
Robinson et al. \cite{Robinson2024ATrial} explored the feasibility of using a humanoid robot to deliver brief mindfulness breathing technique to students in higher education, showing that the robot successfully promoted relaxation, contentment, and engagement, particularly for students experiencing psychological distress. Another study investigated the use of robotic positive psychology coaches at a university campus, where students participated in sessions in their on-campus accommodation. The robot delivered personalised interventions through rapport-building techniques over a series of interactions aimed at improving well-being, mood, and motivation, with the robot effectively improving students' emotional states~\cite{Jeong2022DeployingWell-being}.

The integration of social robots into familiar environments is crucial for maximizing their potential to support emotional well-being. Previous studies found that user evaluations of domestic robots improve with sustained interaction, highlighting that sociability is key to long-term acceptance in domestic environments \cite{Laban2024BuildingTime, laban_ced_2023,DeGraaf2015SharingRobot}. Moreover, studies on robotic well-being coaches in workplaces \cite{spitale_hri_} and in higher education settings \cite{Robinson2021AEducation} highlight the importance of robots adapting to their environments to provide emotional and practical support, demonstrating the versatility of robots in real-world contexts.

\section{Research Questions}
Previous studies have highlighted the potential of robots to support well-being (see Section \ref{related}). 
However, no research has focused 
on how social robots can encourage constructive emotion regulation strategies, 
particularly in real-world settings where they communicate appropriately using LLM. Most interactions aimed at supporting well-being view the robot as a companion (e.g., \cite{Williams2024EnhancingTrial}), rather than an active guide in helping people navigate and manage their emotions. 
Here, we focus on how regular interactions with a robot in familiar environments can support 
and improve emotion regulation. 
Therefore, this study investigates how a robot-mediated cognitive reappraisal intervention 
integrated into familiar settings, can help participants process their emotions more effectively, adopt constructive emotion regulation strategies, and experience emotional upliftment. Beyond demonstrating how robots influence people's well-being, our study seeks to investigate if robots can help individuals change the way they perceive and approach emotionally charged situations, potentially improving their ability to cope in the future. 


Accordingly, we pose three research questions. First, we aim to assess the robot's role in supporting participants' self-regulatory mechanisms of emotion regulation, specifically in understanding emotional experiences and feeling in control. 
Thus, we ask \textbf{(RQ1)}: \textit{To what extent does interacting with a robot influence students' ability to understand and control their emotions?} Next, we assess whether the cognitive reappraisal intervention facilitated by the robot supports the cognitive adaptation of constructive emotion regulation tendencies. Therefore, we ask \textbf{(RQ2)}: \textit{To what extent does interacting with the robot during cognitive reappraisal interventions help students reframe negative emotional experiences and adopt constructive emotion regulation strategies?} 
Beyond addressing the effects of cognitive reappraisal on internal emotional processing, we further explore how repeated interactions with the robot influence participants' outward emotional expressions. Specifically, we examine whether verbal and facial emotional expression change over time and across individual interactions, as behavioural cues of emotion regulation. Thus, we ask \textbf{(RQ3)}: \textit{To what extent does interacting with a robot affect verbal and facial emotional expression over time and throughout each interaction?} This research question aims to determine whether participants become more verbally expressive (e.g., speaking for longer, using more emotion-descriptive words, or displaying changes in the sentiment of their disclosures) and whether their facial expressions systematically shift in arousal or valence during the intervention.
Finally, given the direct benefits of cognitive reappraisal to emotional well-being, 
we evaluate the 
impact of the intervention on participants’ moods and feelings. 
Hence, our final research question is \textbf{(RQ4)}: \textit{To what extent do robot-assisted cognitive reappraisal impact students' well-being and emotional upliftment?}

\section{The Study}


\subsection{Setup}

\begin{figure}[h]
    \centering
    \includegraphics[width=.49\columnwidth]{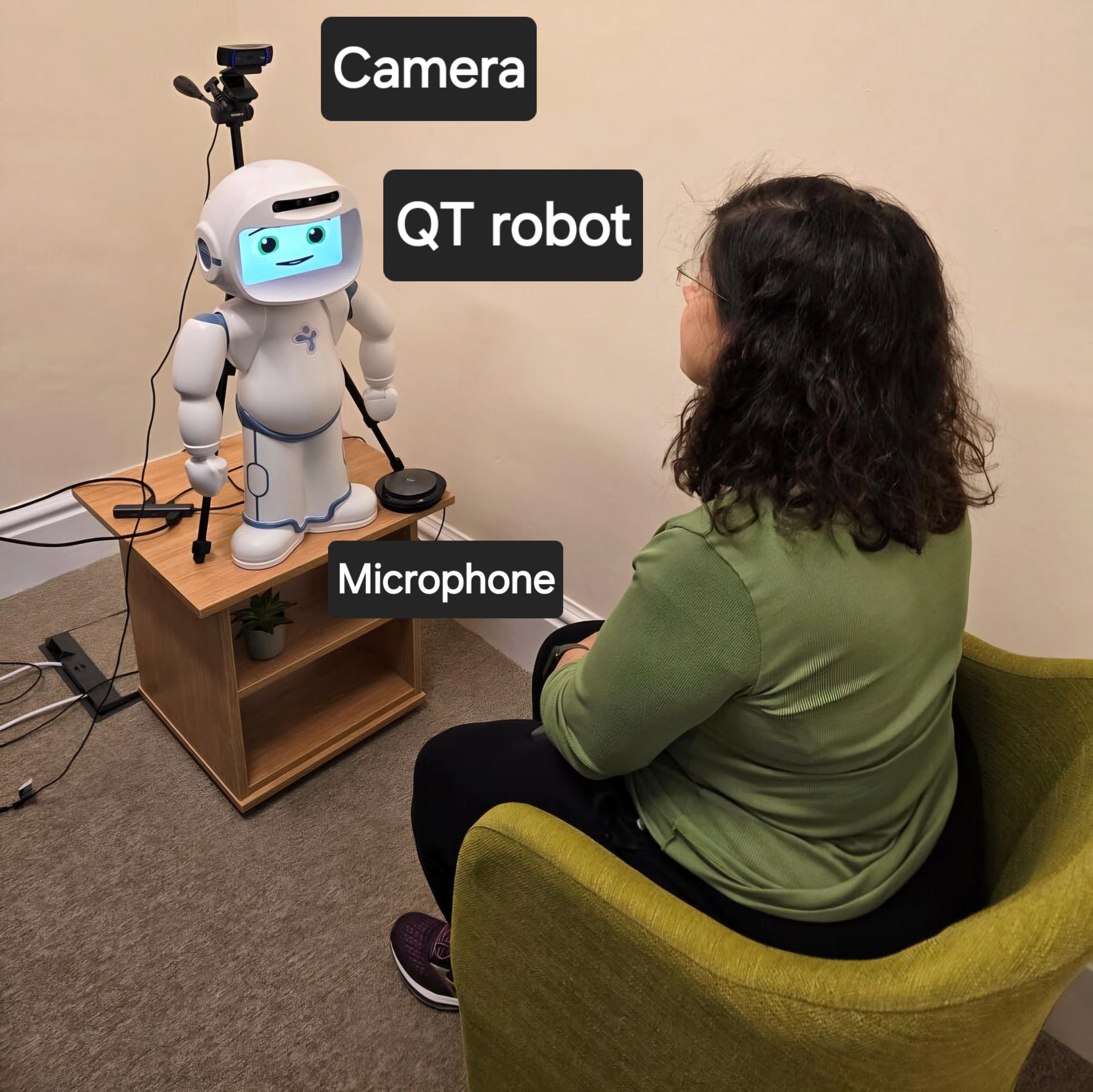} 
    \includegraphics[width=.49\columnwidth]{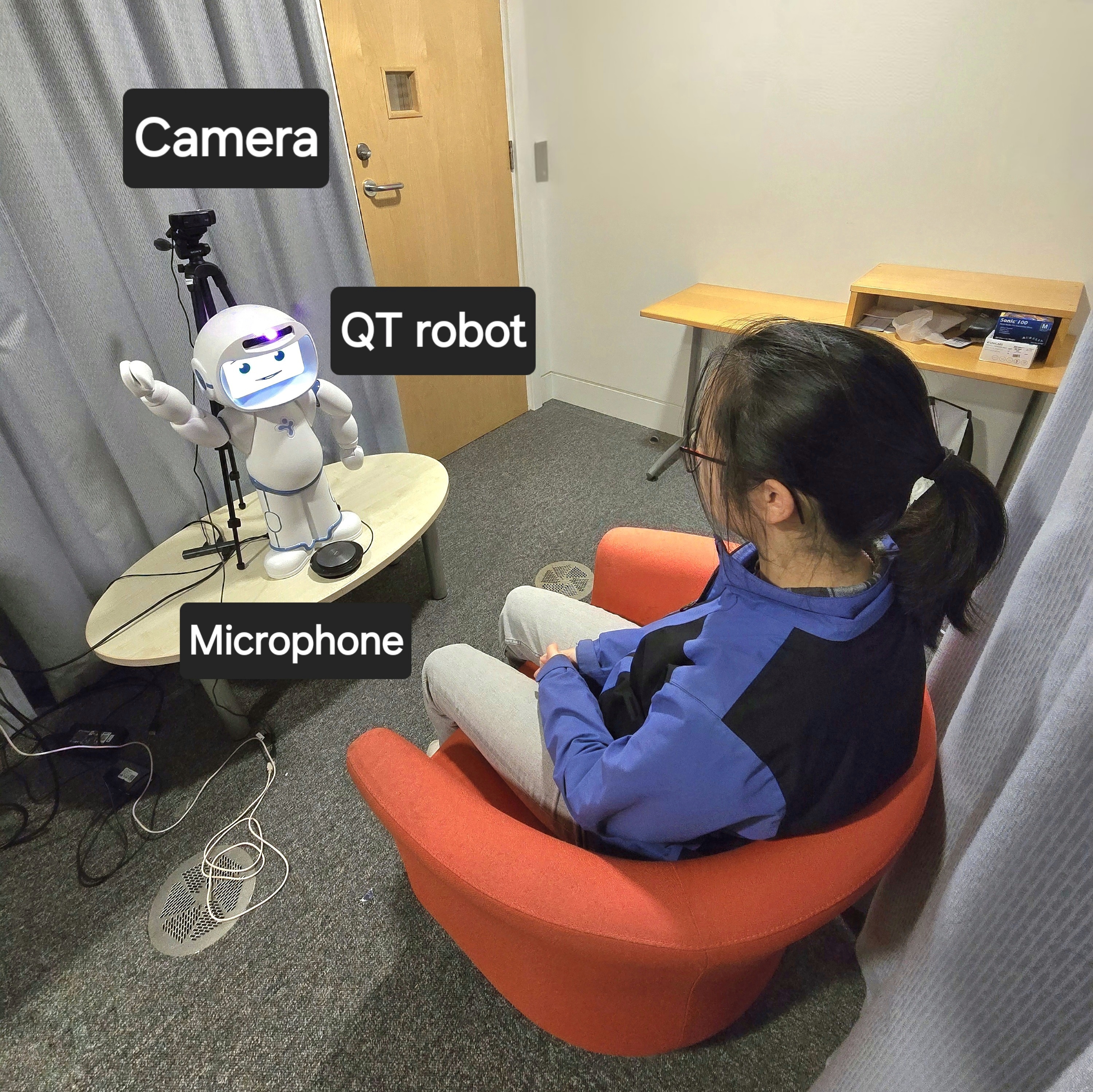}
    \caption{\footnotesize The deployment settings. On the left - C1; On the right - C2.}
    \label{fig:set}
\end{figure}

The study was conducted in two cohorts. In cohort 1 (C1), it took place in a dedicated room within university halls where students live, study, and spend much of their time. In cohort 2 (C2), the study was conducted in a dedicated room in a university department. In this setting as well, the robot was placed in an environment where students work, study, and spend most of their day. In both locations, students attended the sessions as part of their daily schedule within these environments. 
Participants sat on a 
couch positioned in front of 
the robot, which was placed on a low table to match the participants’ eye level. An external video camera was positioned behind the robot, and an external microphone in front of it (see Figure \ref{fig:set}).

\subsection{Participants}
\label{participants}

An \textit{a priori} power analysis was conducted for a linear mixed-effects model \cite{RefWorks:410} with a single group with 5 repeated measurements per participant to determine the required sample size. The analysis assumed an effect size of $f$ = .25, an $\alpha$ error probability of .05, and a desired power (1 - $\beta$) of .80. 
The correlation among repeated measures was set at .5, and 
$\epsilon$ was assumed to be 1. The power analysis indicated that a sample size of 21 participants would be sufficient to achieve the desired power. The analysis yielded a 
$\lambda$ of 13.13, a critical F-value of 2.49, and 4 numerator and 80 denominator degrees of freedom. The actual power achieved 
was .82. 

Accordingly, we recruited 21 participants on a voluntary basis. Participants were recruited in two cohorts: one cohort of students from university halls  (C1) and an additional cohort of students from a university department (C2). Recruitment for C1 was facilitated through posters displayed around the hall's premises and the hall’s social media channel on Instagram. Recruitment for C2 was facilitated through a single email call for participants. In both cohorts, students attended the sessions as part of their daily schedule within these environments 
Participants ranged in age from 18 to 27 (M = 20.62, $\pm$2.66), with 43\% identifying as female, and more than 50\% being native English speakers. The University 
requires a minimum score of 7.5 on the IELTS exam, 
ensuring that all participants possess a sufficient level of English proficiency. 
More than 70\% of participants were undergraduate students. 
42.11\% of the participants were from STEM departments
, 36.84\% from the social sciences
, and 21.05\% from the humanities. 

\subsection{Robotic platform \& System}

We selected the QTrobot (referred to as QT) by LuxAI S.p.A as the robotic platform to deliver the intervention. QT features a screen face, an RGB-D camera, and a microphone array. It is a 90 cm tall, child-like robot with a 2 DOF neck, 4 DOF arms (including shoulders, elbows, and hands), and static legs. We selected the QT for this study as it offers an approachable humanoid interface that encourages conversational interactions without eliciting uncanny perceptions (as demonstrated by previous studies using QT, e.g., \cite{Dennler2023AArm,Spitale2024AppropriatenessEvaluation,Axelsson2024RobotsRecommendations,Spitale2022SociallyEmpathy,Spitale2021ComposingInteraction}). Additionally, it allows for smooth integration of the LLM with the system's components, ensuring precise voice synchronisation and efficient user input processing. 
We used an open-source system \cite{Spitale2023VITA:Coaching} (with turned-off adaptation functionality to avoid confounding effects) to develop and implement the LLM-powered (GPT 3.5) intervention described in Section \ref{exp_d} and Figure \ref{flowchart}. We used AWS Polly’s Amy synthesised voice and synchronised the robot’s mouth movements with the spoken voice using Amazon Polly visemes. Participants' speech was captured using Google Cloud Speech-to-Text and submitted directly to the OpenAI API. The robot’s motors were only used to wave hello and goodbye.

\subsection{Ethics and Safety}

The study was approved by the department's ethics committee, and all participants signed an informed consent and were debriefed after their participation. To ensure that the LLM-powered interactions follow appropriate safety guidelines, the robotic system used in this study incorporates safety measures such as an integrated content moderation layer to filter inappropriate inputs, rigorous adversarial testing to validate response robustness, and prompt engineering techniques to constrain the topic and tone, all embedded within a structured conversational framework and guided by comprehensive ethical oversight (see \cite{Spitale2023VITA:Coaching} for details on the system's safety measures). To test these measures, we simulated various scenarios and conducted pilots with the university hall’s well-being team prior to the study, ensuring that the system was performing in line with their expectations.

\subsection{Experimental Design and Intervention} 
\label{exp_d}

We conducted a long-term study using a single-condition, repeated-measures experimental design. The study was approved by the departmental ethics committee. Participants interacted with the robot QT over five sessions within two weeks. The intervention was based on the PERMA model \cite{Seligman2018PERMAWell-being,Seligman2011Flourish:Well-being.b}, a positive psychology framework designed to capture the multidimensional aspects of 
well-being: \textbf{P}ositive Emotions (session 1), \textbf{E}ngagement (session 2), \textbf{R}elationships (session 3), \textbf{M}eaning in Life (session 4), and \textbf{A}ccomplishments (session 5). PERMA model is 
a valuable framework for  structuring well-being interventions by addressing key dimensions of positive psychology
\cite{Kovich2023ApplicationStudents,Kern2015AFramework}. The goal of the intervention was to elicit cognitive reappraisal, 
helping participants to change their subjective interpretation of emotional stimuli \cite{Gross1998Antecedent-Physiology}. 

\begin{figure}[h!]
\centering

\noindent\resizebox{\columnwidth}{!}{
\begin{tikzpicture}[node distance=1.5cm]

\node (start) [block_start_end] {\textbf{Start}};

\node (positive) [block2, below of=start] {\textbf{Positive-Connotation Question (PCQn)}};

\node (followup1) [block, below of=positive] {\textbf{Follow-up (Fn)}};

\node (recognition1) [block, below of=followup1] {\textbf{Recognition (Rn)}};

\node (negative) [block3, right of=recognition1, xshift=4cm] {\textbf{Negative-Connotation Question (NCQ)}};

\node (reappraisal) [block, above of=negative] {\textbf{Reappraisal (CR)}};

\node (followup2) [block, above of=reappraisal] {\textbf{Follow-up (F3)}};


\node (end) [block_start_end, above of=followup2] {\textbf{End}};

\draw [arrow] (start) -- (positive);
\draw [arrow] (positive) -- (followup1);
\draw [arrow] (followup1) -- (recognition1);
\draw [arrow] (recognition1) -- (negative);
\draw [arrow] (negative) -- (reappraisal);
\draw [arrow] (reappraisal) -- (followup2);
\draw [arrow] (followup2) -- (end);

\draw [dashed, ultra thick, orange, -{Latex[length=3mm]}] (positive.east) -- (reappraisal.west);

\draw [dashed, ultra thick, orange, -{Latex[length=3mm]}] (followup1.east) -- (reappraisal.west);

\draw [dashed, ultra thick, -{Latex[length=3mm]}] (recognition1.west) -- ++(-2,0) node[midway, above] {\textbf{X 2}} |- (positive.west);

\end{tikzpicture}
}

\caption{\footnotesize The interaction flow. 
The blue block indicates pre-scripted Positive-Connotation Questions (PCQn), while the Red block represents the pre-scripted Negative-Connotation Question (NCQ). Green blocks involve responses powered by the LLM - Follow-up (Fn), Recognition (Rn), and Reappraisal (CR). The split arrow from Rn to PCQn suggests repetition (x2). The orange split arrow shows that CR is based on PCQn and Fn responses.}
\label{flowchart}

\end{figure}
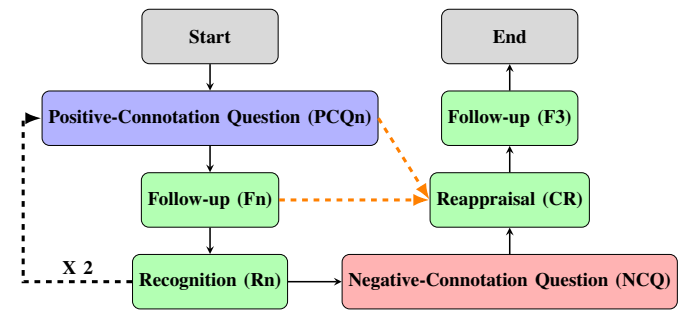

To achieve this, QT facilitated the intervention in a structured manner (see Figure \ref{flowchart}). Each interaction began with pre-scripted greetings, followed by QT asking a pre-scripted, positive connotation question (\textbf{PCQn}) about one of the PERMA topics, corresponding to the session number. After the participant responded, their response was submitted to GPT-3.5 with the prompt, ``\textit{Follow up on what I just shared with you}" (\textbf{Fn}). After the participant responded again, all of their responses were resubmitted to GPT-3.5 with the prompt, ``\textit{Recognise the feelings and emotions I shared with you and ask me if I am ready for another question}" (\textbf{Rn}). In line with previous studies examining LLMs' 
ability to recognise and appraise emotions from disclosures \cite{Tak2023IsEmotion,Tak2024GPT-4Perspective,Rathje2024GPTAnalysis}, this segment of the interaction was designed to facilitate emotional reflection and summarise the interaction before proceeding. Further, this process (PCQn to Rn, see Figure \ref{flowchart}) occurred twice, with questions framed around the same PERMA topic. 


Following this procedure 
QT asked the participant one pre-scripted, negative connotation question (\textbf{NCQ}) about the session's PERMA topic, encouraging the participant to reflect on a challenge they might be experiencing in a central area of their life. 
Rather than following up directly on the participant’s answer, QT aimed to guide them through reappraising their negative experiences or emotions by reintroducing relevant elements from their earlier responses to the positive connotation questions. To facilitate this, the participant's response was submitted to GPT-3.5 with the prompt 
``\textit{Help me to see the positive aspects of the negative experience I shared with you, and assist me in cognitive reappraisal by highlighting the positive elements from my previous responses. Then, ask me a question that will encourage self-reflection}" (\textbf{CR}). After the participant responded, their answer was submitted again to GPT-3.5 with the prompt, ``\textit{Follow up on what I just shared with you}" (\textbf{F3}), to engage them further in discussing their reappraised feelings. 
After the participant responded, the robot ended the session with a pre-scripted message, thanking the participant and wishing them well. See supplementary material 1 for all interaction items.

Before starting the experiment, we conducted a short validation pilot with 5 students and 2 well-being advisors from the university halls. Each student and well-being advisor participated in one session. After the session, we asked them about their experience, specifically focusing on how QT 
reappraised their negative disclosures. All validation pilot participants confirmed that QT successfully reappraised their disclosures, and the 
advisors confirmed that QT's technique aligned with their expectations for robotics student support in the halls.

\subsection{Procedure \& Measures}
\label{procedure}

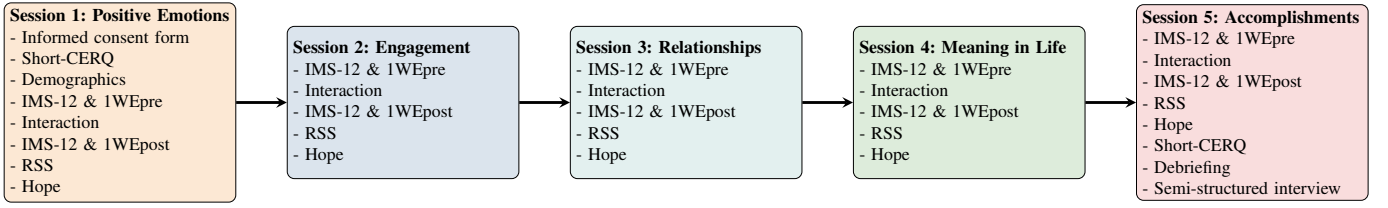
\begin{figure*}[h!]
\centering

\noindent\resizebox{\textwidth}{!}{
\begin{tikzpicture}[
    block/.style = {rectangle, draw, rounded corners, text width=4.33cm, minimum height=3cm, align=left, fill=#1, text opacity=1, fill opacity=0.2},
    arrow/.style = {ultra thick, ->, >=stealth},
    ]

    \definecolor{session1}{HTML}{F28E2C}
    \definecolor{session2}{HTML}{4E79A7}
    \definecolor{session3}{HTML}{76B7B2}
    \definecolor{session4}{HTML}{59A14F}
    \definecolor{session5}{HTML}{E15759}

    \node[block=session1] (session1) {
        \textbf{Session 1: Positive Emotions} \\
        - Informed consent form \\
        - Short-CERQ \\
        - Demographics \\
        - IMS-12 \& 1WEpre \\
        - Interaction\\
        - IMS-12 \& 1WEpost \\
        - RSS \\
        - Hope
    };

    \node[block=session2, right=of session1] (session2) {
        \textbf{Session 2: Engagement} \\
        - IMS-12 \& 1WEpre \\
        - Interaction\\
        - IMS-12 \& 1WEpost \\
        - RSS \\
        - Hope
    };

    \node[block=session3, right=of session2] (session3) {
        \textbf{Session 3: Relationships} \\
        - IMS-12 \& 1WEpre \\
        - Interaction\\
        - IMS-12 \& 1WEpost \\
        - RSS \\
        - Hope
    };

    \node[block=session4, right=of session3] (session4) {
        \textbf{Session 4: Meaning in Life} \\
        - IMS-12 \& 1WEpre \\
        - Interaction\\
        - IMS-12 \& 1WEpost \\
        - RSS \\
        - Hope
    };

    \node[block=session5, right=of session4] (session5) {
        \textbf{Session 5: Accomplishments} \\
        - IMS-12 \& 1WEpre \\
        - Interaction\\
        - IMS-12 \& 1WEpost \\
        - RSS \\
        - Hope \\
        - Short-CERQ \\
        - Debriefing \\
        - Semi-structured interview
    };

    \draw[arrow] (session1.east) -- (session2.west);
    \draw[arrow] (session2.east) -- (session3.west);
    \draw[arrow] (session3.east) -- (session4.west);
    \draw[arrow] (session4.east) -- (session5.west);

\end{tikzpicture}}
\caption{\footnotesize Flow of the long-term study according to the PERMA model and the questionnaires reported in each session.}
\label{fig:experiment-flowchart}
\end{figure*}

Students scheduled their participation 
(first session) using a Calendly 
page 
with information about the study. 
Upon attending the first session, participants were given further instructions and explanations about the study through an informed consent form, which they were required to sign before participating.  
Before interacting with QT for the first time, participants completed the Short Cognitive Emotion Regulation Scale (Short-CERQ \cite{Garnefski2006CognitiveCERQ-short}; M = 3.38, $\pm$.48, $\alpha$ = .83)  reporting their 
emotion regulation tendencies, and also 
several demographic values (age, identified gender, biological sex, nationality, first language).

Before starting each session, participants reported their mood using the 12-item Immediate Mood Scaler (IMS-12 \cite{Nahum2017}; M = 4.72, $\pm$.93, $\alpha$ = .92) and described their emotions and feelings in one word using a text box (1WEpre). Before the first interaction, participants were given basic instructions about their interactions with QT and its functionalities. Once participants indicated they were ready to begin, the experimenter left the room to give them privacy to interact with QT as they wished. Each interaction lasted about 17 minutes on average ($\pm$ 268 seconds). After completing the interaction, participants again reported their mood using the IMS-12 \cite{Nahum2017} (M = 5.35, $\pm$.89, $\alpha$ = .94) and described their emotions and feelings in one word using a text box (1WEpost). Following each interaction, participants completed the Regulatory Effectiveness of Social Support Questionnaire (RSS) \cite{Zee2020RegulatorySupport.} to assess the extent to which the interaction impacted their understanding of their emotions (3 items; M = 4.67, $\pm$1.37, $\alpha$ = .91) and feelings of control over their emotions (3 items; M = 4.33, $\pm$1.27, $\alpha$ = .93). In each session participants also reported their feelings of hope after interacting with QT 
\cite{Herth1992AbbreviatedEvaluation} (7 items, M = 3.99, $\pm$.65, $\alpha$ = .83).

In the final session, participants reported their emotion regulation tendencies using the Short-CERQ \cite{Garnefski2006CognitiveCERQ-short} after completing the interaction. Upon finishing each session, participants 
scheduled their next session, 
were thanked for their participation and provided with contact information of the researchers and 
of 
support services, in case needed. Furthermore, after the final session, participants were debriefed about their participation. 
Upon completing the fifth session, we  interviewed 13 participants that agreed to participate in a semi-structured interview.  
See Figure \ref{fig:experiment-flowchart} for the study's flow. The interactions were recorded (audio and video), and interaction logs were collected to extract measures (see Section \ref{data_ana}) of speech duration (M = 43.01, $\pm$27.38), compound sentiment (M = .17, $\pm$.19), adjective count (M = 5.42, $\pm$4.84), facial valence (M = -.01, $\pm$.83), and facial arousal (M = .26, $\pm$.29).

\subsection{Data Analysis}
\label{data_ana}

Linear mixed-effects models (LME) were employed using  lme4 for R \cite{Bates2015FittingLme4} to evaluate repeated measures across the five sessions, accounting for both fixed effects (the session number, and pre/post measures) and random effects (accounting for participant variance). These models were selected for their ability to manage 
repeated-measures data and ensure robustness in capturing individual differences \cite{Bell2019FixedChoice}. Additionally, they were well-suited for data from two cohorts, as modeling participants as random effects inherently accounts for individual variability within and across cohorts, ensuring that cohort-level differences do not confound the overall analysis \cite{Hesser2015ModelingInterventions}. Significance was calculated using lmerTest \cite{Kuznetsova2017LmerTestModels} applying Satterthwaite’s method \cite{Satterthwaite1946AnComponents}. 
To evaluate how participants' feelings changed due to the intervention, 
sentiment was extracted from participants' responses to 1WEpre and 1WEpost (see Section \ref{procedure}) using TextBlob \cite{Loria2024Textblob0.18.0.post0}, a 
tool that calculates sentiment polarity ranging from -1 (extremely negative) to +1 (extremely positive). 

The video and audio data were analysed to extract affective meaning. Speech data were processed using Parselmouth \cite{RefWorks:473}, a python-based off-the-shelf tool for PRAAT analysis \cite{RefWorks:474} to identify speech duration in seconds. The interaction logs were analysed for verbal expression measures, including sentiment and adjective count. Sentiment was extracted from participants' disclosures using TextBlob \cite{Loria2024Textblob0.18.0.post0}. To analyse the distribution of adjectives from participants' disclosures we applied Part-of-Speech (POS) tagging using the Natural Language Toolkit (NLTK; \cite{Bird2009NaturalPython}). All text entries were processed using NLTK’s word tokenization method. POS tagging was performed on each tokenized sentence to identify adjectives, defined as words classified as JJ (adjective, e.g., happy), JJR (comparative adjective, e.g., happier), and JJS (superlative adjective, e.g., happiest) following Penn Treebank POS tags (see \cite{Marcus1993BuildingTreebank}). The count of adjectives per text entry was computed accordingly. For evaluating participants' facial expressions during the intervention, the interaction videos were processed using the FaceChannel dimensional model \cite{Barros2020TheRecognition}, an off-the-shelf package that predicts valence (how positive or negative the facial expression is) and arousal (the intensity of the facial expression) values \cite{Barrett1998DiscreteFocus}, ranging from -1 to 1 (see \cite{Barros2020TheRecognition}).

Qualitative feedback from 13 participants was gathered through semi-structured interviews after competing their fifth session. See Supplementary Material 2 for all interview questions. A thematic analysis \cite{Nowell2017ThematicCriteria} was conducted to extract recurring themes from participants' descriptions of their experiences during the intervention. Two trained human evaluators analysed participants’ answers, identifying key themes using a systematic, iterative coding process. Initially, they familiarised themselves with the data, then generated preliminary codes that captured both the manifest and latent content of the responses. These codes were subsequently clustered into broader themes through an inductive approach, ensuring that the emerging categories accurately reflected the nuances in the data  \cite{Nowell2017ThematicCriteria}. The analysis focused on identifying patterns in 
reflection and introspection, self-regulation, and the perceived affective value of interacting with the robot. Key themes emerged around the participants' sense of emotional support, empowerment, and cognitive reappraisal facilitated by the robot. These qualitative insights were used to complement 
the understanding of the quantitative results, providing a more holistic view of the intervention’s impact.  

\section{Results}

\subsection{Improved Self-Regulatory Mechanisms}

\begin{figure}[h]
    \centering
    \includegraphics[width=.49\columnwidth]{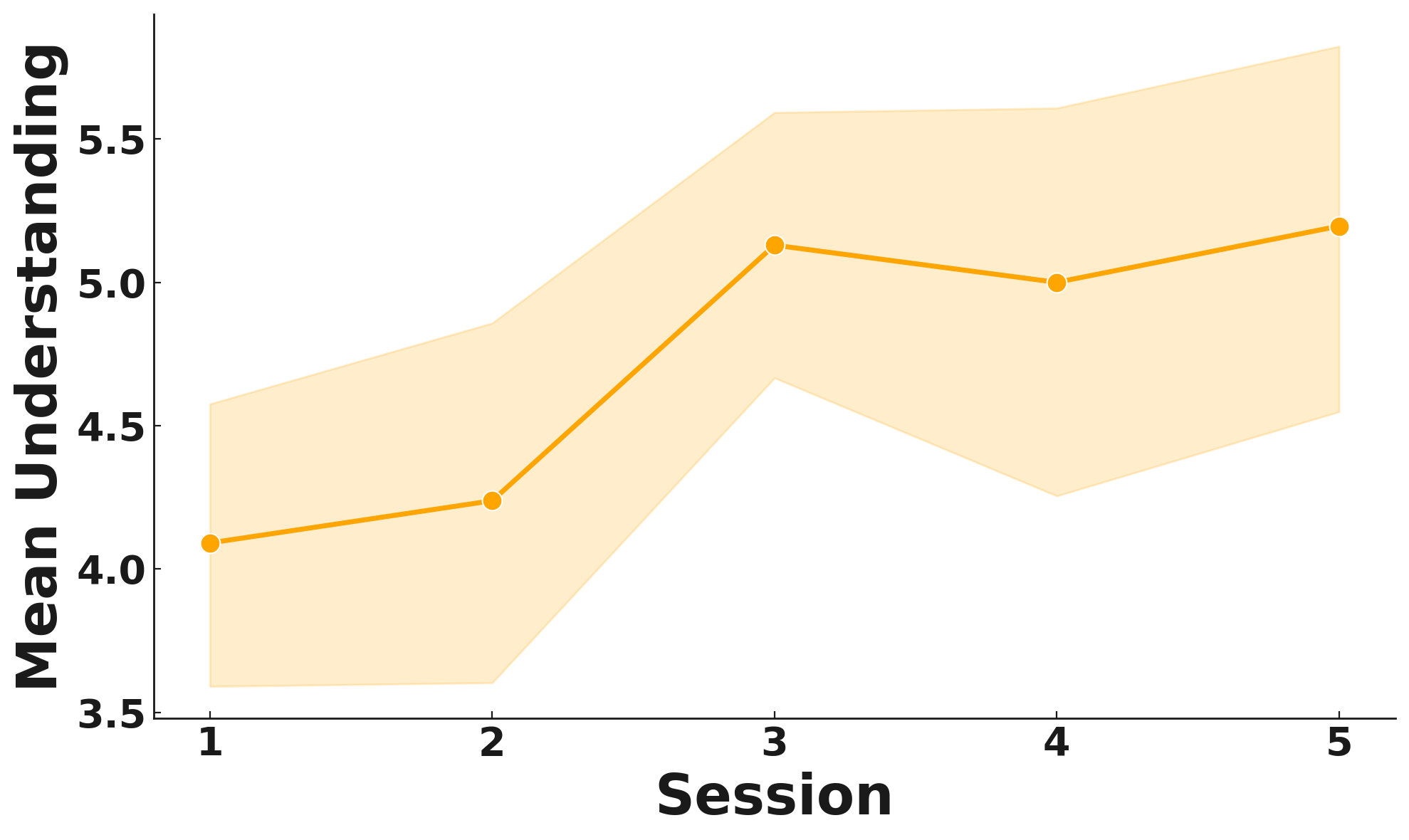}
    \includegraphics[width=.49\columnwidth]{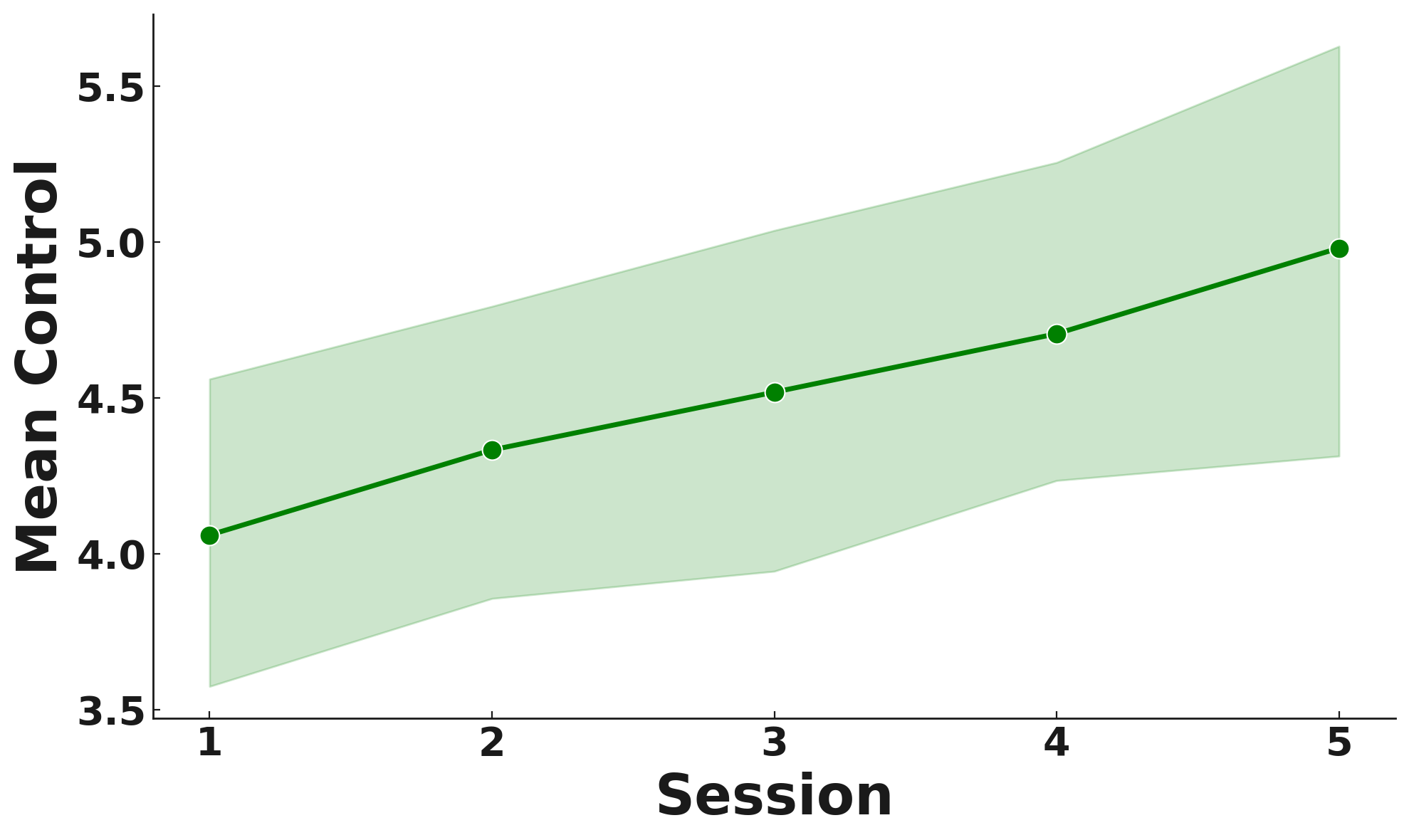}
    \caption{\footnotesize From left to right: (1) 
    mean score of QT's support in understanding by session. (2) mean score of QT's support in control by session.
    }
    \label{fig:rss}
\end{figure}

The intervention 
had a significant positive effect on participants’ self-regulatory mechanisms, as indicated by improvements in understanding of their emotional situations and sense of control over the five sessions. As demonstrated in Figure \ref{fig:rss}, the results stress that despite the variance between the participants ($\pm$1), participants’ understanding of their emotional situations increased significantly over time, 
$\beta$ = .30, $SE = .06$, 
$p$ $<$ .001 (see Table \ref{rss_t}). 
Participants' sense of control over their emotions also showed a significant increase, $\beta$ = .24, $SE = .06$, 
$p$ $<$ .001 (see Table \ref{rss_t} and Figure \ref{fig:rss}), despite the variance across participants ($\pm$.90). 

\begin{table}[htbp!]
\caption{\small Mixed Effects Results of Understanding \& Control}\label{rss_t}
\centering\resizebox{\columnwidth}{!}{%
\begin{tabular}{lcccc}
\toprule
 & \multicolumn{2}{c}{Understanding the Situation} & \multicolumn{2}{c}{Feeling in Control} \\
\cmidrule(lr){2-3} \cmidrule(lr){4-5}
\textbf{Predictors} & \textbf{Estimates} & \textbf{95\%CI} & \textbf{Estimates} & \textbf{95\%CI} \\
\midrule
Intercept & 3.78*** & 3.19–4.36 & 3.79*** & 3.24–4.34 \\
Session & 0.30*** & 0.17–0.42 & 0.24*** & 0.11–0.36 \\
\midrule
\multicolumn{5}{l}{\textbf{Random Effects}} \\
$\sigma^2$ & \multicolumn{2}{c}{0.73} & \multicolumn{2}{c}{0.68} \\
$\tau_{00}$ & \multicolumn{2}{c}{1.00} & \multicolumn{2}{c}{0.80} \\
ICC & \multicolumn{2}{c}{0.58} & \multicolumn{2}{c}{0.54} \\
N & \multicolumn{2}{c}{21} & \multicolumn{2}{c}{21} \\
Observations & \multicolumn{2}{c}{94} & \multicolumn{2}{c}{94} \\
Marginal $R^2$ / Conditional $R^2$ & \multicolumn{2}{c}{0.093 / 0.617} & \multicolumn{2}{c}{0.071 / 0.573} \\
\midrule
\multicolumn{5}{l}{\footnotesize{* $p$$<$0.05 ** $p$$<$0.01 *** $p$$<$0.001}} \\
\bottomrule
\end{tabular}}
\end{table}

The qualitative feedback from the interviews reveals how QT facilitated participants' understanding of their emotions and empowered them to feel more in control of their emotional challenges. The ability to confront and understand emotional challenges 
was a recurring theme. P03 remarked, ``\textit{When it} [QT] \textit{asked me these sorts of questions, I think I should no longer avoid it, and it’s the time to act}" acknowledging that QT prompted them to address long-avoided emotional 
issues. This 
was echoed by P09, who stated, ``\textit{QT would pick on one of the things that I said that stood out more… and then they would give a question asking me to elaborate… which kind of forced me to think more deeply and closely about that subject that I wouldn't normally do}" reflecting how QT helped them confront and take control of their stressors in a more thoughtful way. Many participants felt that QT’s responses encouraged them to reflect on aspects of their emotional lives they typically avoided. For example, P12 shared, ``\textit{QT makes you really reflect on stuff that you don't usually engage in those conversations in everyday}" highlighting QT’s role in promoting introspection. Similarly, P13 noted, “\textit{they touch on some subjects that I know I normally have a hard time with, but it }[QT] \textit{didn’t bother me that much. It }[QT] \textit{was asking questions, and I felt everything was really delicate and within reason – it }[QT] \textit{didn’t push too much}”, reflecting how QT balanced delicate emotional topics with thoughtful engagement, allowing participants to address difficult emotions without feeling overwhelmed. P03 remarked, “\textit{Talking to it is like I am getting a secondhand option, a different perspective that I should also consider, and it is really hard to see}”, showing how QT offered a new perspectives. 

Participants also highlighted how QT’s supportive comments fostered a sense of control over their emotional states. As P10 noted, 
``\textit{QT helps organize my thoughts and speak about it in a more organized manner}". QT’s ability to help participants confront issues they typically avoided was also evident in P09’s experience: “\textit{I think, in my opinion, like it's really nice in the moment, and then maybe might calm you down. But ultimately, the academic stress is still there, the social stress is still there. You haven't really acknowledged or confronted it, whereas with QT you get the opportunity to confront it}”. Moreover, QT served as a unique emotional resource compared to other forms of emotional support. As P06 noted, “\textit{I think therapy dogs act more as a distraction, whereas with QT, if you want it to be a distraction, it can be a distraction, but also, if you want it could be a space to discuss the actual problem and try to work through the actual problem. QT is a resource for you}”. This reflects how QT provided participants with both a supportive space for emotional regulation and a tool to actively address their challenges, rather than simply offering comfort or distraction.


\subsection{Cognitive Changes in Emotion Regulation Tendencies} 

\begin{table*}[htbp!]
\centering
\caption{\small Mixed Effects Results for Cognitive Changes in Emotion Regulation}\label{er_table}
\resizebox{.75\textwidth}{!}{%
\begin{tabular}{lcccccccccc}
\toprule
 & \multicolumn{2}{c}{Acceptance} & \multicolumn{2}{c}{Positive Reappraisal} & \multicolumn{2}{c}{Putting-Into-Perspective} & \multicolumn{2}{c}{Positive Refocusing} \\
\cmidrule(lr){2-3} \cmidrule(lr){4-5} \cmidrule(lr){6-7} \cmidrule(lr){8-9}
\textbf{Predictors} & \textbf{Estimates} & \textbf{95\%CI} & \textbf{Estimates} & \textbf{95\%CI} & \textbf{Estimates} & \textbf{95\%CI} & \textbf{Estimates} & \textbf{95\%CI} \\
\midrule
Intercept & 3.47*** & 2.99–3.95 & 3.89*** & 3.48–4.29 & 3.38*** & 2.97–3.78 & 2.78*** & 2.34–3.21 \\
Session & 0.13* & 0.02–0.23 & 0.14* & 0.03–0.26 & 0.15** & 0.05–0.25 & 0.12* & 0.03–0.22 \\
\midrule
\multicolumn{9}{l}{\textbf{Random Effects}} \\
$\sigma^2$ & \multicolumn{2}{c}{\centering 0.43} & \multicolumn{2}{c}{\centering 0.50} & \multicolumn{2}{c}{\centering 0.33} & \multicolumn{2}{c}{\centering 0.33} \\
$\tau_{00}$ & \multicolumn{2}{c}{\centering 0.53} & \multicolumn{2}{c}{\centering 0.13} & \multicolumn{2}{c}{\centering 0.35} & \multicolumn{2}{c}{\centering 0.47} \\
ICC & \multicolumn{2}{c}{\centering 0.55} & \multicolumn{2}{c}{\centering 0.21} & \multicolumn{2}{c}{\centering 0.51} & \multicolumn{2}{c}{\centering 0.58} \\
N & \multicolumn{2}{c}{\centering 21} & \multicolumn{2}{c}{\centering 21} & \multicolumn{2}{c}{\centering 21} & \multicolumn{2}{c}{\centering 21} \\
Observations & \multicolumn{2}{c}{\centering 42} & \multicolumn{2}{c}{\centering 42} & \multicolumn{2}{c}{\centering 42} & \multicolumn{2}{c}{\centering 42} \\
Marginal $R^2$ / Conditional $R^2$ & \multicolumn{2}{c}{\centering 0.061 / 0.578} & \multicolumn{2}{c}{\centering 0.111 / 0.300} & \multicolumn{2}{c}{\centering 0.116 / 0.566} & \multicolumn{2}{c}{\centering 0.070 / 0.614} \\
\midrule
\multicolumn{9}{l}{\footnotesize{* $p$$<$0.05 ** $p$$<$0.01 *** $p$$<$0.001}} \\
\bottomrule
\end{tabular}}
\end{table*}

The intervention 
had a significant positive impact on participants' constructive emotion regulation strategies (see Table \ref{er_table}). 
The results stress that despite the variance between the participants ($\pm$.37), participants’ use of positive reappraisal significantly increased following the intervention, $\beta$= .14, $SE = .06$, 
$p = .017$. 
A significant increase was also observed in acceptance, $\beta$ = .13, $SE = .06$,  
$p = .025$, 
despite the variance between the participants ($\pm$.73). 
Participants also reported a significant improvement in positive refocusing, $\beta$ = .12, $SE = .05$, 
$p = .017$, 
regardless of the variance between the participants ($\pm$.68). Also, despite the variance between the participants ($\pm$.59), a significant increase was found in putting into perspective, $\beta$ = .15, $SE = .05$, 
$p = .005$.

The qualitative 
results reflect the effectiveness of QT in promoting cognitive reappraisal. Participants frequently mentioned a shift in how they perceived and responded to challenging situations. 
For instance, P04 noted, ``\textit{Stressful events felt way more significant to me... but now, I place more emphasis on the positive things}", demonstrating a reappraisal of negative events by placing greater value on positive experiences. Similarly, P08 highlights this reappraisal process: ``\textit{Normally in life we focus on just the negative – something small that happens every day and we put all our attention there. We think that we’re worthless the way we have such short amounts. It's nice to remember that whatever you have done are your accomplishments, you know you are not worthless}", suggesting that participants learned to view their actions from a more balanced perspective, reducing the emotional impact of everyday stressors and enhancing their focus on personal achievements. Many participants described how QT 
helped them reframe negative emotions and gain insight into their situations. For instance, P13 shared, ``\textit{It was nice to see a positive outlook on the fact that I’m occupying myself. I thought about it as a negative thing because I was lonely and bored, and the robot said it was a good thing. It was nice to hear it}", 
Similarly, P07 explained, ``\textit{It helped me get some perspective on the situation. It helped me feel a bit encouraged that maybe I could change things and do things in a different way in the future}" showing how QT enabled them to see potential solutions to their emotional difficulties.

Participants also described how QT facilitated a sense of acceptance and self-compassion, which further supports the use of reappraisal. P02 expressed, ``\textit{It really helped me be rather accepting of myself... the world doesn’t end because I didn’t do that one thing that day}", indicating that they had learned to reframe moments of perceived failure as part of a broader, more forgiving perspective on their efforts and accomplishments. 
P03 shared, ``\textit{Talking to QT made me feel confident about different aspects of my past}", reflecting a reappraisal of previously self-critical judgements. Similarly, P06 emphasised that QT guided participants towards ``\textit{discovering more things about yourself}", P06 added: ``\textit{It's not necessarily QT Telling you and giving you a new perspective, but it's under QT's guidance}". 
Even when QT was not directly providing new insights, merely speaking to QT and answering to QT's questions supported the practice of reappraisal. This suggests that QT fostered an environment for participants to internally reframe their emotional narratives. 
Interestingly, the self-reflection facilitated by QT was often described as occurring unconsciously, 
highlighting the subtle yet impactful nature of the intervention. As P09 noted, ``\textit{The self-reflection that I experienced was also kind of subconscious, 
it wasn't intentional. It just 
happened while I was interacting with QT}". 

The intervention showed no significant effects on maladaptive emotion regulation strategies. 
Participants' use of rumination did not significantly change, 
$\beta$ = -.03, $SE = .04$, 
$p = .362$. 
Similarly, no significant change was found in self-blame, $\beta$ = .01, $SE = .05$, 
$p = .801$. 
The intervention also had no significant effect on catastrophizing, $\beta$ = .01, $SE = .04$, 
$p = .711$, or blaming others, $\beta$ = -.03, $SE = .05$, 
$p = .550$. 


\subsection{The effects of the intervention on Verbal Expression}

\begin{table}[h!]
\caption{\small Mixed Effects Results for Verbal Expression}
\label{tab:verb}
\centering
\resizebox{\columnwidth}{!}{%
\begin{tabular}{lccc}
\toprule
 & \multicolumn{1}{c}{Speech Duration} & \multicolumn{1}{c}{Sentiment Polarity} & \multicolumn{1}{c}{Adjective Use} \\
\cmidrule(lr){2-2} \cmidrule(lr){3-3} \cmidrule(lr){4-4}
\textbf{Predictors} & \textbf{Estimates 95\% CI} & \textbf{Estimates 95\% CI} & \textbf{Estimates 95\% CI} \\
\midrule
Intercept & 29.24$^{***}$ 20.36--38.12 & 0.23$^{***}$ 0.19--0.28 & 3.57$^{***}$ 2.11--5.04 \\
Session & 3.30$^{***}$ 2.02--4.57 & -0.01 -0.01--0.01 & 0.48$^{***}$ 0.24--0.73 \\
Interaction Order & 0.93$^{*}$ 0.07--1.79 & -0.01$^{***}$ -0.02---0.01 & 0.10 -0.06--0.27 \\
\midrule
\multicolumn{4}{l}{\textbf{Random Effects and Model Statistics}} \\
$\sigma^2$ & 474.30 & 0.035 & 17.48 \\
$\tau_{00}$ & 281.33 & 0.01 & 6.19 \\
ICC & 0.37 & 0.04 & 0.26 \\
N & 21 & 21 & 21 \\
Observations & 619 & 619 & 619 \\
Marginal $R^2$ / Conditional $R^2$ & 0.257 / 0.534 & 0.093 / 0.617 & 0.132 / 0.364 \\
\midrule
\multicolumn{4}{l}{\footnotesize{* $p$$<$0.05 ** $p$$<$0.01 *** $p$$<$0.001}} \\
\bottomrule
\end{tabular}%
}
\end{table}

\begin{figure}[h!]
    \centering
    \includegraphics[width=.49\columnwidth]{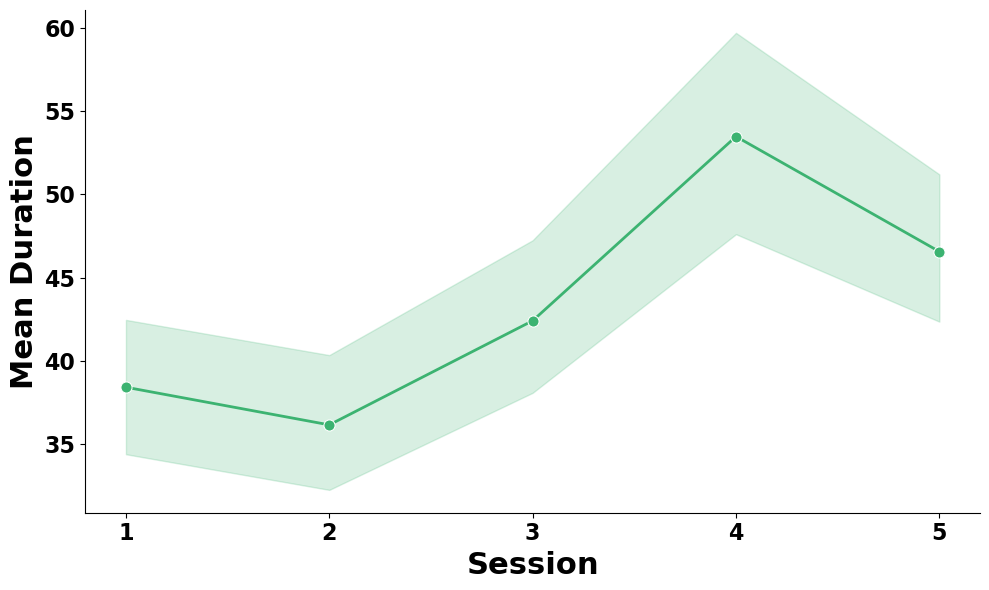}
    \includegraphics[width=.49\columnwidth]{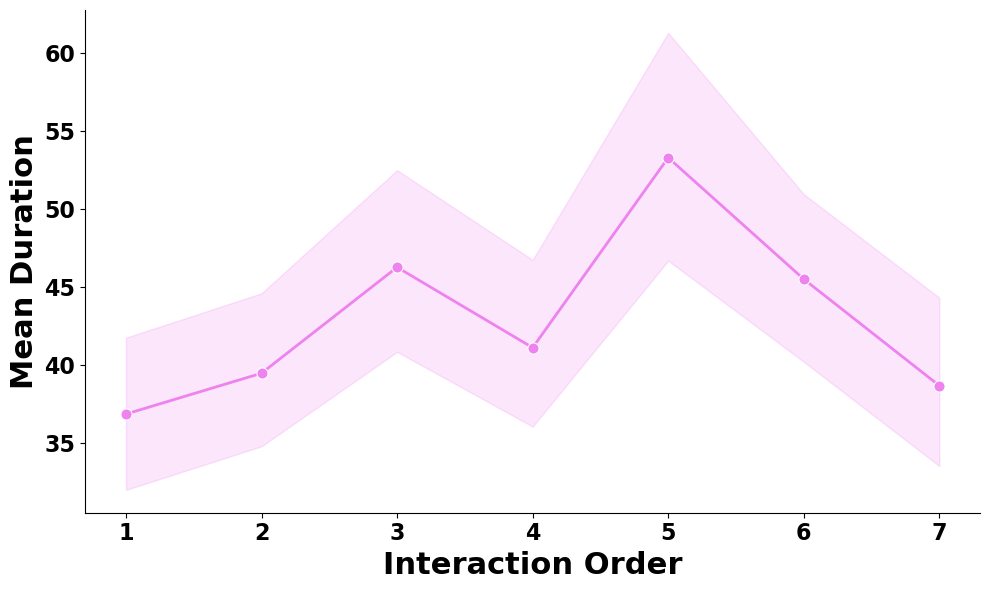}
    \caption{\footnotesize From left to right: (1) 
    mean speech duration by session. (2) mean speech duration by the disclosure item.
    }
    \label{fig:dur}
\end{figure}

The progression of the intervention sessions, as well as the order of disclosures within each interaction, had a significant positive effect on the duration of participants' speech disclosures. Despite individual variability between participants (\(\pm 16.77\)), the duration of participants' speech disclosures showed a significant increase over time across sessions, \(\beta = 3.30\), \(SE = 0.65\), \(p < .001\) (see Table \ref{tab:verb} and Figure \ref{fig:dur}). Similarly, as demonstrated in Figure \ref{fig:dur}, the duration of participants' speech disclosures significantly increased within each disclosure as the interaction progressed, \(\beta = 0.93\), \(SE = 0.44\), \(p = .034\), despite variance across participants (\(\pm 16.78\); see Table \ref{tab:verb}). However, when observing the trend in Figure \ref{fig:dur}, we identified a wave-like pattern in interaction duration that corresponds to the structure of the intervention (see Figure \ref{flowchart}). When modelling the effect of disclosure order within interactions using a Generalized Additive Mixed Model (GAMM), a nonlinear relationship between the order of disclosures and speech duration emerged. Unlike the mixed-effects model, which assumed a linear effect of session progression, the GAMM employed a cubic B-spline to flexibly estimate changes in interaction duration across different points in the interaction sequence, following the unique structure of the intervention. This approach allows for detecting potential fluctuations in interaction duration that may not follow a strictly increasing or decreasing pattern \cite{Vatter2015GeneralizedStructures,Rigby2005GeneralizedShape}, but would predict duration changes based on the reappraisal task. The smoothed effect of disclosure order revealed significant increases in interaction duration at multiple points during interactions. Specifically, duration was significantly higher at the second, fourth, and sixth splines, \(\beta = 47.11\), \(SE = 17.75\), \(p = .008\); \(\beta = 62.19\), \(SE = 17.82\), \(p < .001\), and \(\beta = 1.82\), \(SE = 4.07\), \(p = .655\), respectively. In contrast, duration significantly decreased at the first, third, and fifth splines, \(\beta = -23.16\), \(SE = 12.99\), \(p = .075\); \(\beta = -50.28\), \(SE = 22.72\), \(p = .027\); and \(\beta = -22.26\), \(SE = 12.99\), \(p = .087\), respectively.

We also found a significant negative effect of the order of disclosures within each interaction on sentiment polarity of participants' disclosures' content.
Despite individual variation ($\pm$0.01), sentiment scores exhibited a negative significant effect over the order of participants' disclosures within each interaction, $\beta$ = -0.01, $SE$ = 0.01, $p$ $<$ .001. Conversely, session number did not significantly effect the disclosures' sentiment, $\beta$ = -0.01, $SE$ = 0.01, $p$ = .575, implying that sentiment changes were more closely tied to the sequential order within interactions rather than to progression throughout the duration of the intervention (see Table \ref{tab:verb}).

\begin{figure}[h!]
    \centering
    \includegraphics[width=.49\columnwidth]{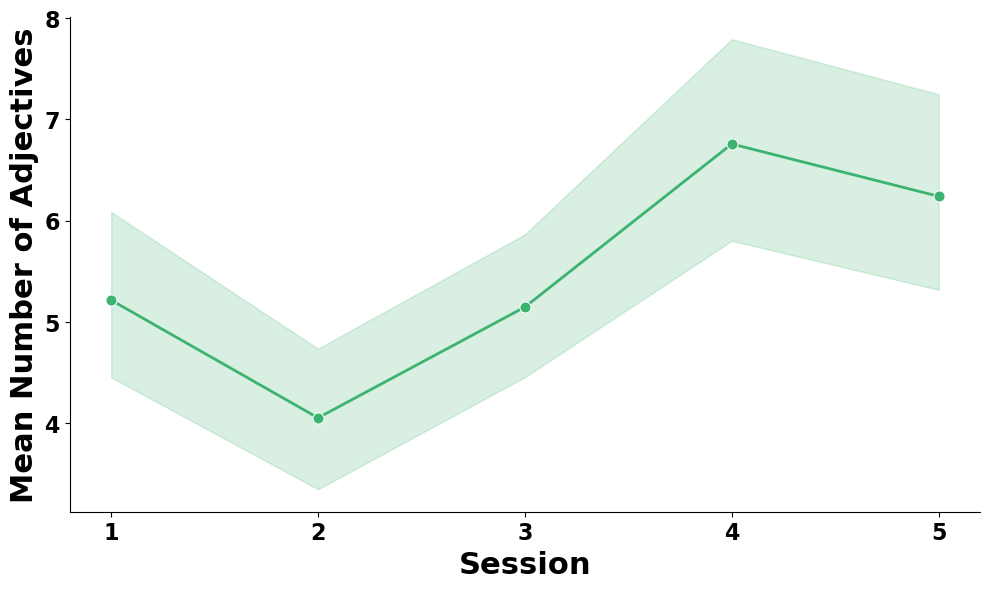}
    \includegraphics[width=.49\columnwidth]{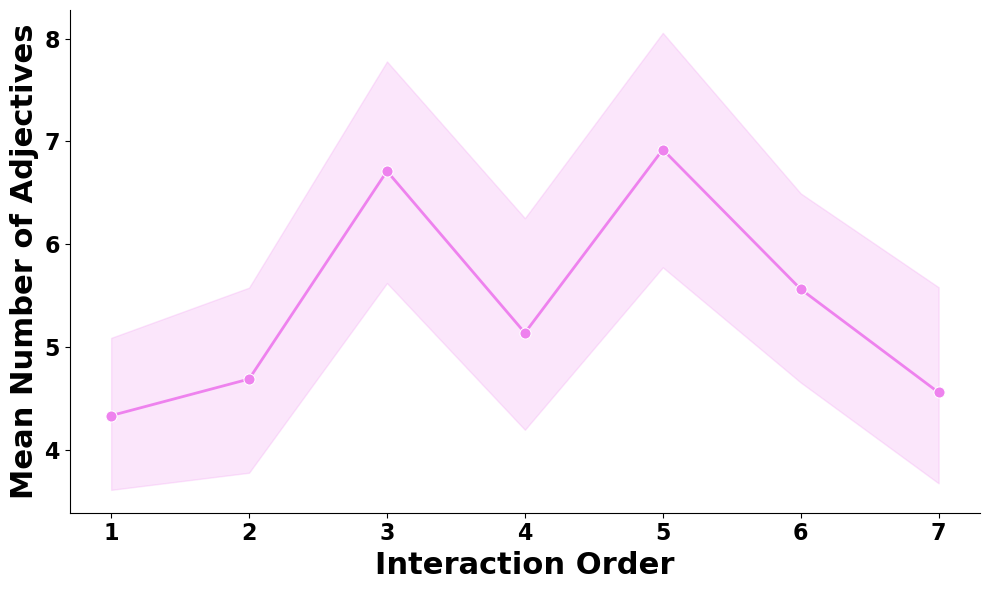}
    \caption{\footnotesize From left to right: (1) 
    mean number of adjectives by session. (2) mean number of adjectives by the disclosure item.
    }
    \label{fig:adj}
\end{figure}

Finally, the number of adjectives used by participants significantly increased across sessions, while the order of disclosures within each interaction did not show a significant effect. Despite individual variability between participants (\(\pm 6.19\)), the number of adjectives used showed a significant increase over time across sessions, \(\beta = 0.48\), \(SE = 0.13\), \(p < .001\) (see Table \ref{tab:verb} and Figure \ref{fig:adj}). The order of disclosures within each interaction did not significantly effect the number of adjectives used in linear terms, \(\beta = 0.10\), \(SE = 0.08\), \(p = .222\). 


However, when observing the trend in Figure \ref{fig:adj}, we identified a trend in adjective use that corresponds to the structure of the intervention (see Figure \ref{flowchart}). When modelling the effect of disclosure order within interactions using a GAMM, a nonlinear relationship between the disclosures order within each interaction and adjective use emerged. Unlike the mixed-effects model, which assumed a linear effect of session progression, the GAM employed a cubic B-spline to flexibly estimate changes in adjective use across different points in the interaction sequence, following the unique structure of the intervention. This approach allows for detecting potential fluctuations in adjective use that may not follow a strictly increasing or decreasing pattern \cite{Vatter2015GeneralizedStructures,Rigby2005GeneralizedShape}, but would predict their usage according to the reappraisal task. The smoothed effect of disclosure order revealed significant increases in adjective use at multiple points during interactions. Specifically, adjective use was significantly higher at the second, fourth, and sixth splines, \(\beta = 9.50\), \(SE = 1.67\), \(p < .001\); \(\beta = 9.51\), \(SE = 1.68\), \(p < .001\), and \(\beta = 4.50\), \(SE = 0.54\), \(p < .001\), respectively. In contrast, adjective use did not significantly change at the first, third and fifth splines, \(\beta = -0.015\), \(SE = 1.91\), \(\beta = 2.87\), \(SE = 1.61\), \(p = .074\), \(p = .994\) and \(\beta = 2.13\), \(SE = 1.93\), \(p = .270\), respectively.

\subsection{The effects of intervention on facial expression}

~\begin{table}[htbp!]
\caption{\small Mixed Effects Results of Arousal and Valence}\label{arousal_valence_t}
\centering\resizebox{\columnwidth}{!}{%
\begin{tabular}{lcccc}
\toprule
\textbf{Predictors} & \multicolumn{2}{c}{\textbf{Arousal}} & \multicolumn{2}{c}{\textbf{Valence}} \\
\cmidrule(lr){2-3} \cmidrule(lr){4-5} 
& \textbf{Estimates} & \textbf{95\% CI} & \textbf{Estimates} & \textbf{95\% CI} \\
\midrule
Intercept & 0.15$^{***}$ & 0.08 -- 0.23 & 0.34$^{***}$ & 0.13 -- 0.48 \\
Session Number & 0.03$^{***}$ & 0.03 -- 0.03 & -0.02$^{*}$ & -0.03 -- -0.01 \\
Interaction Order & 0.01$^{***}$ & 0.00 -- 0.01 & -0.05$^{***}$ & -0.06 -- -0.05 \\
\midrule
\multicolumn{5}{l}{\textbf{Random Effects}} \\
$\sigma^2$ & \multicolumn{2}{c}{0.06} & \multicolumn{2}{c}{0.52} \\
$\tau_{00}$ (Participant) & \multicolumn{2}{c}{0.03} & \multicolumn{2}{c}{0.15} \\
ICC & \multicolumn{2}{c}{0.34} & \multicolumn{2}{c}{0.22} \\
N (Participants) & \multicolumn{2}{c}{21} & \multicolumn{2}{c}{21} \\
Observations & \multicolumn{2}{c}{8436} & \multicolumn{2}{c}{8436} \\
Marginal $R^2$ / Conditional $R^2$ & \multicolumn{2}{c}{0.10 / 0.42} & \multicolumn{2}{c}{0.162 / 0.475} \\
\midrule
\multicolumn{5}{l}{\footnotesize{* $p$$<$0.05, ** $p$$<$0.01, *** $p$$<$0.001}} \\
\bottomrule
\end{tabular}}
\end{table}

\begin{figure}[h!]
    \centering
    \includegraphics[width=.49\columnwidth]{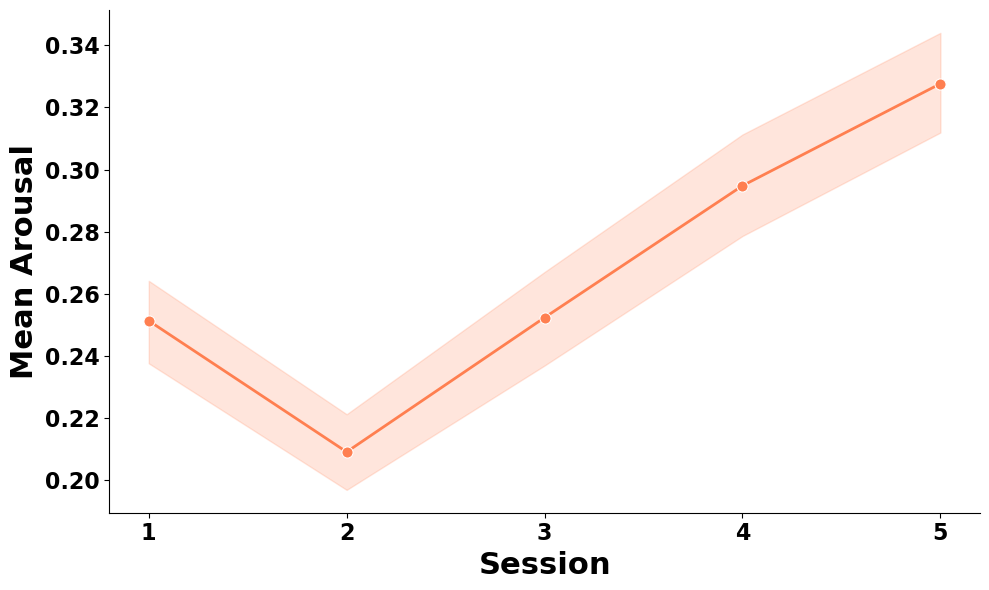}
    \includegraphics[width=.49\columnwidth]{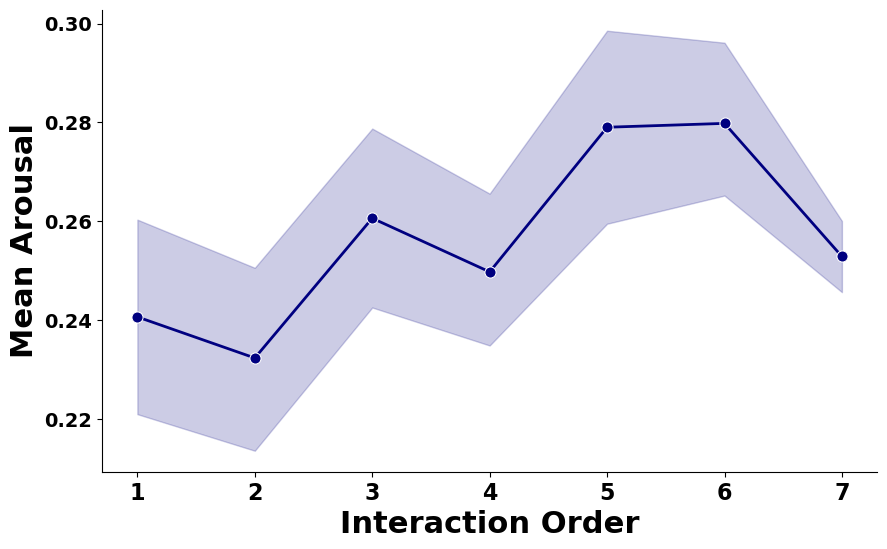}
    \caption{\footnotesize From left to right: (1) 
    mean score of arousal by session. (2) mean score of arousal by the disclosure item.
    }
    \label{fig:aro}
\end{figure}

The progression of the intervention sessions, as well as the order of disclosures within each interaction, had a significant positive effect on participants' facial arousal levels. Despite individual variability between participants (\(\pm 0.03\)), arousal showed a significant increase over time across sessions, \(\beta = 0.030\), \(SE = 0.002\), \(p < .001\) (see Table \ref{arousal_valence_t} and Figure \ref{fig:aro}). As demonstrated in Figure \ref{fig:aro}, arousal significantly increased within each interaction as the interaction progressed, \(\beta = 0.006\), \(SE = 0.001\), \(p < .001\), despite variance across participants (\(\pm 0.03\); see Table \ref{arousal_valence_t}).

~\begin{figure}[h!]
    \centering
    \includegraphics[width=.49\columnwidth]{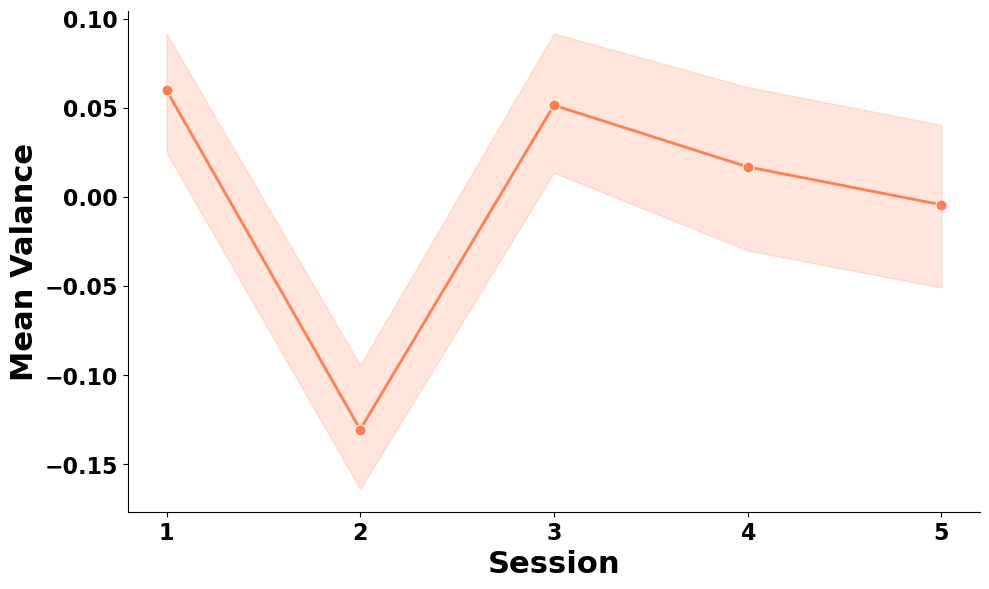}
    \includegraphics[width=.49\columnwidth]{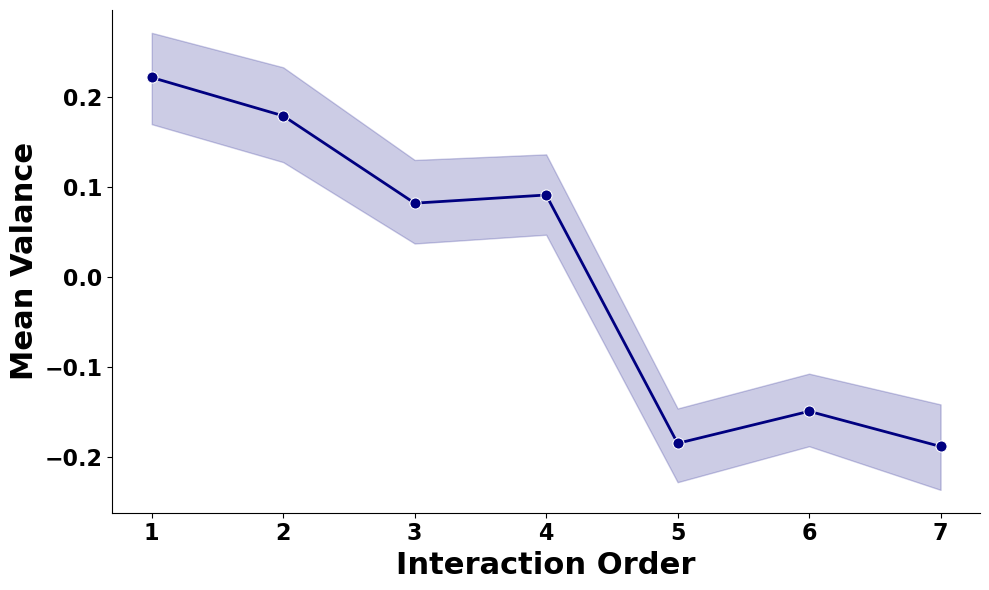}
    \caption{\footnotesize From left to right: (1) 
    mean score of valance by session. (2) mean score of valance by the disclosure item.
    }
    \label{fig:val}
\end{figure}

The progression of the intervention sessions, as well as the order of disclosures within each interaction, had a significant negative effect on participants' facial valence levels, indicating that when participants were engaged in negative disclosure and reappraisal, their valence decreased. Despite individual variability between participants ($\pm 0.01$), valence showed a significant decline within each interaction as the interaction progressed, $\beta = -0.05$, $SE = 0.01$, $p < .001$ (see Table \ref{arousal_valence_t} and Figure \ref{fig:val}). Session progression had a significant negative effect on valence, $\beta = -0.02$, $SE = 0.01$, $p = .012$, despite variance across participants ($\pm 0.01$; see Table \ref{arousal_valence_t} and Figure~\ref{fig:val}).

\subsection{Influence on Well-being}

\begin{figure}[hb!]
    \centering
    \includegraphics[width=.7\columnwidth]{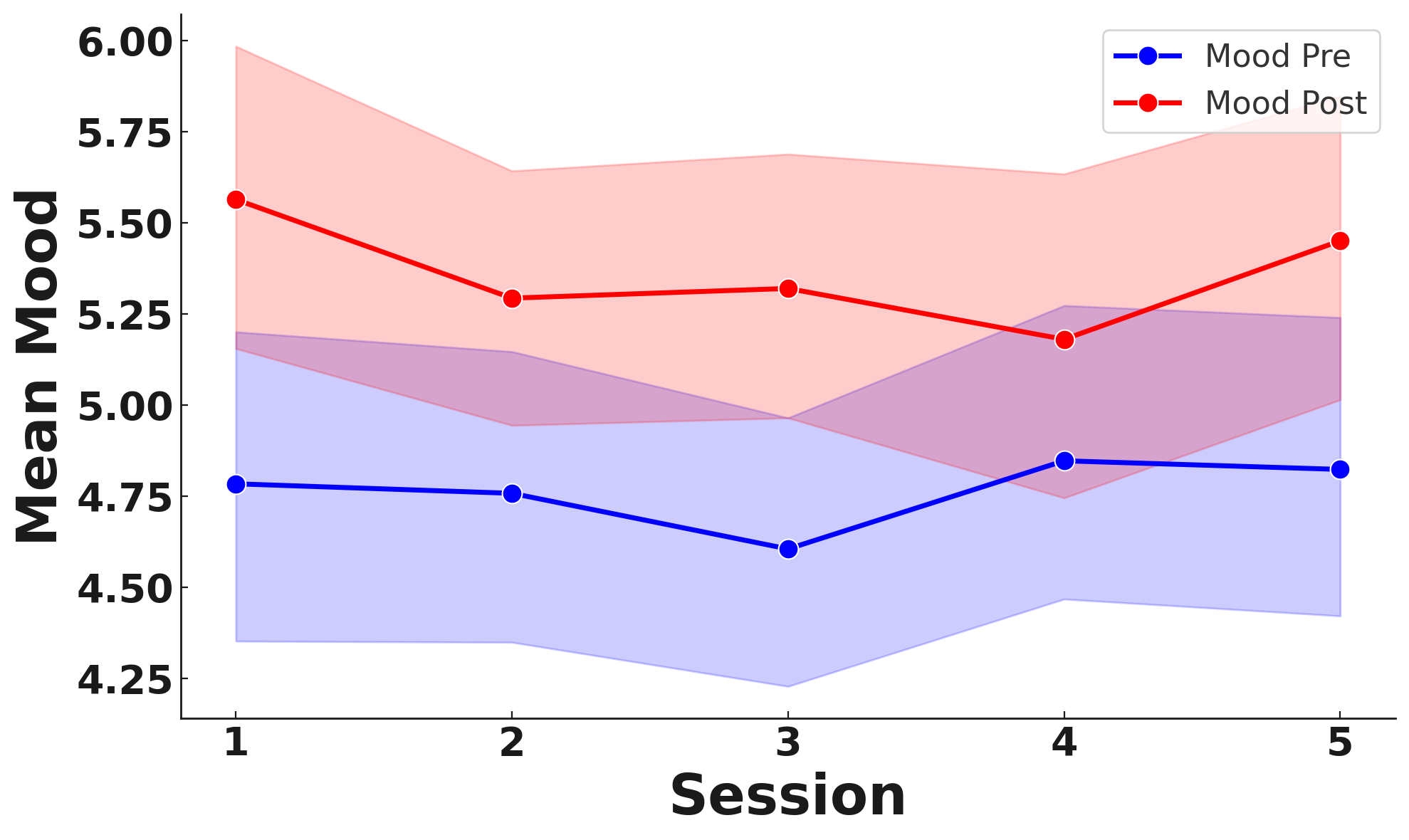}
    \caption{\footnotesize Mean mood scores before and after the interaction by session number}
    \label{fig:ims}
\end{figure}

As demonstrated in Figure \ref{fig:ims}, the results stress that despite the variance between the participants ($\pm$.62), the intervention 
had a significant positive effect on participants’ mood change following each session. Results revealed a significant improvement in mood after each session, with a fixed effect estimate of $\beta$ = .59, $SE = .10$, 
$p$ $<$ .001. 
In contrast, there was no significant effect on mood change over time, $\beta$ = .02, $SE = .04$, 
$p = .564$ (see Table \ref{wb_t}). 

\begin{figure}[hb!]
    \centering
    \includegraphics[width=.47\columnwidth]{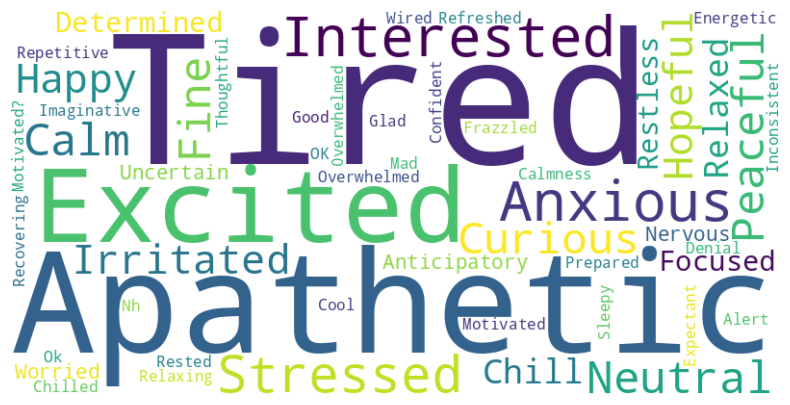}
    \includegraphics[width=.47\columnwidth]{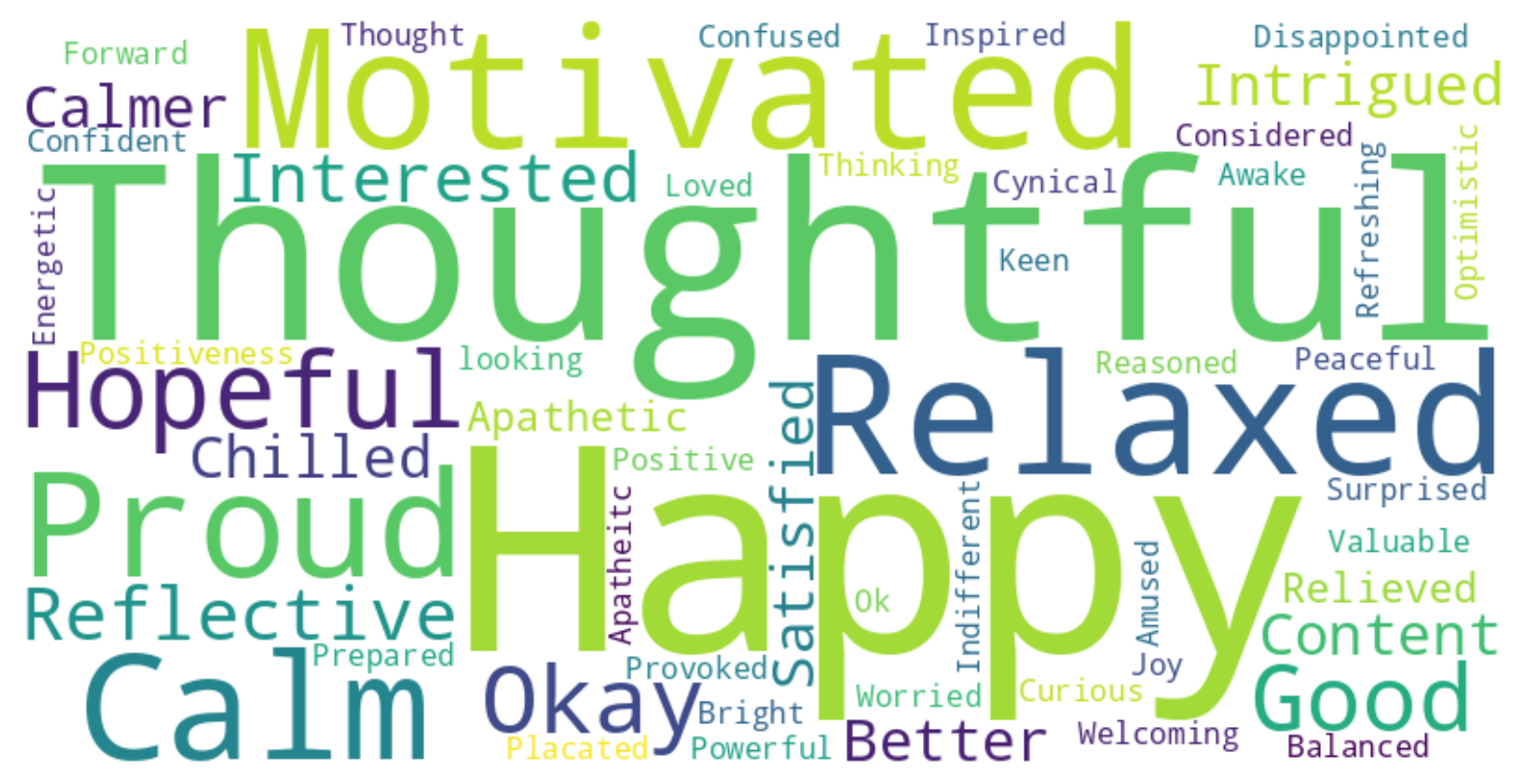}
    \caption{\footnotesize Word clouds illustrating the sentiment distribution before (left) and after (right) the interaction.}
    \label{fig:wordcloud}
\end{figure}

The most frequent words describing participants' emotions and feelings before their participation (``1WEpre"; see Figure \ref{fig:wordcloud}) were ``\textit{Tired}" (9), ``\textit{Apathetic}" (6), ``\textit{Excited}" (6), ``\textit{Interested}" (6), ``\textit{Anxious}" (5), ``\textit{Stressed}" (4), and ``\textit{Fine}" (3). The most frequent words describing participants' emotions and feelings after their participation (``1WEpost"; see Figure \ref{fig:wordcloud}) were ``\textit{Happy}" (6), ``\textit{Motivated}" (5), ``\textit{Calm}" (5), ``\textit{Thoughtful}" (5), ``\textit{Proud}" (5), ``\textit{Relaxed}" (5), and ``\textit{Hopeful}" (4). 

Despite the variance between the participants ($\pm$.78), the intervention 
had a significant positive effect on improvement in sentiment after each session, with a fixed effect estimate of $\beta = .16$, $SE = .04$, 
$p < .001$ (see Table \ref{wb_t}). 
There was no significant effect of session number on sentiment change over time, $\beta = -.02$, $SE = .02$, 
$p = .313$ (see Table \ref{wb_t}). 
Moreover, 
the results stress that despite the variance across participants ($\pm$.53), feelings of hope grew gradually over time, with a significant fixed effect by session, $\beta$ = .06, $SE = .03$, 
$p = .048$ (see Table \ref{wb_t}). 

~

The qualitative feedback from the interviews underscores the emotional benefits of interacting with QT. 
P12 expressed how the opportunity to talk about emotions provided a valuable outlet, saying, ``\textit{It was nice to take time out of your day to speak to someone about like how you're feeling, which I don't do}". This 
was echoed by P06
: ``\textit{QT will usually respond to my response with 
an additional question or maybe a comment on 
feeling happy for me, proud of me, or changing my perspective on saying negative or something like that, I found those comments really meaningful}". 
P01 reflected on how engaging with QT helped them relax, stating, ``\textit{I think it was probably in the second session that I warmed up to it. It was quite relaxing}". Similarly, P05 expressed 
``\textit{I feel better after that than I did before, which surprised me, even with drawbacks that they} [QT] \textit{identified}. \textit{I guess the idea is that there's an anonymity and a lack of judgment from it, which is good and helpful}". P13 also highlighted the positive emotional effect of the interaction, saying, ``\textit{It was really fun, just gaining new perspectives, and it filled me with optimism}". P10 highlighted the importance of such emotional outlets, stating, “\textit{We should have these outlets for people within 
halls when probably they are feeling the most stressed and wanting to speak to someone}”. 
P09 noted a surprising benefit, saying, “\textit{I never realized that I would feel a lot better after talking about my experiences, whether it be positive or negative. Just saying my experience out loud was really helpful in ways that I did not expect}". This suggests that the simple act of affect labeling (see \cite{Torre2018PuttingRegulation}) with QT provided emotional relief, even for those who did not initially anticipate it. P11 also described how QT’s interactions helped them reframe their thoughts and feel more relaxed: ``\textit{I think it reframed it in like a sort of positive way to make you think about it again. It made me feel more relaxed}”. 

~\begin{table}[h!]
\centering
\caption{\small Mixed Effects Results of Mood, Sentiment, and Hope}\label{wb_t}
\resizebox{\columnwidth}{!}{%
\begin{tabular}{lcccccc}
\toprule
 & \multicolumn{2}{c}{Mood} & \multicolumn{2}{c}{Sentiment} & \multicolumn{2}{c}{Hope} \\
\cmidrule(lr){2-3} \cmidrule(lr){4-5} \cmidrule(lr){6-7}
\textbf{Predictors} & \textbf{Estimates} & \textbf{95\%CI} & \textbf{Estimates} & \textbf{95\%CI} & \textbf{Estimates} & \textbf{95\%CI} \\
\midrule
Intercept & 4.78*** & 4.42–5.13 & 0.27*** & 0.15–0.39 & 3.85*** & 3.56–4.14 \\
Mood Change (pre/post) & 0.59*** & 0.40–0.78 & 0.16*** & 0.08–0.24 & -- & -- \\
Session & 0.02 & -0.05–0.09 & -0.02 & -0.04–0.01 & 0.06* & 0.00–0.11 \\
\midrule
\multicolumn{6}{l}{\textbf{Random Effects}} \\
$\sigma^2$ & \multicolumn{2}{c}{0.46} & \multicolumn{2}{c}{0.08} & \multicolumn{2}{c}{0.14} \\
$\tau_{00}$ & \multicolumn{2}{c}{0.38} & \multicolumn{2}{c}{0.02} & \multicolumn{2}{c}{0.28} \\
ICC & \multicolumn{2}{c}{0.45} & \multicolumn{2}{c}{0.24} & \multicolumn{2}{c}{0.67} \\
N & \multicolumn{2}{c}{21} & \multicolumn{2}{c}{21} & \multicolumn{2}{c}{21} \\
Observations & \multicolumn{2}{c}{188} & \multicolumn{2}{c}{188} & \multicolumn{2}{c}{94} \\
Marginal $R^2$ / Conditional $R^2$ & \multicolumn{2}{c}{0.095 / 0.503} & \multicolumn{2}{c}{0.037 / 0.562} & \multicolumn{2}{c}{0.016 / 0.671} \\
\midrule
\multicolumn{6}{l}{\footnotesize{* $p$$<$0.05 ** $p$$<$0.01 *** $p$$<$0.001}} \\
\bottomrule
\end{tabular}%
}
\end{table}

\section{Discussion}

\subsection{Impact on Emotion Regulation}
The significant increases in participants’ understanding of their emotional situations and their sense of control over the five sessions suggest that the intervention strengthened self-regulatory mechanisms \textbf{(RQ1)}. These improvements are crucial for effective emotion regulation and have been linked to better psychological outcomes \cite{Gross2015EmotionProspects}. The enhancement of constructive cognitive emotion regulation strategies 
indicates that participants were not only experiencing immediate emotional relief but were also developing long-term skills to manage their emotions more effectively \textbf{(RQ2)}. Interestingly, the absence of significant changes in maladaptive emotion regulation strategies, such as rumination and self-blame, suggests that while participants adopted more constructive strategies, they did not increase their reliance on maladaptive ones. This is particularly noteworthy as it indicates that the intervention does not inadvertently promote negative coping mechanisms. Despite the cognitive effort required to adopt constructive strategies—and the intuitive ease of falling back on maladaptive ones—participants were able to enhance their use of positive emotion regulation without exacerbating maladaptive tendencies. This balance highlights the intervention’s potential in fostering healthy emotional processing without unintended side effects. Qualitative feedback further supports these findings, with participants reporting that the robot facilitated deeper introspection and offered a supportive space for discussing emotions. Participants highlighted the robot’s role in helping them confront  
emotional challenges, providing new perspectives and promoting self-reflection. The robot’s non-judgemental and consistent presence may have contributed to these outcomes, aligning with previous research suggesting that social robots support 
emotional expression \cite{Laban2024SharingFeel}. 

\subsection{Effects on Expression}

Throughout the intervention, participants expressed themselves more richly and openly, using longer speech durations and more descriptive emotional adjectives, while also being more expressive non-verbally (i.e., through facial expressions) \textbf{(RQ3)}. This key finding reinforces previous HRI research, showing that people gradually open up to robots when disclosing about themselves over time \cite{Laban2024BuildingTime,laban_ced_2023, Laban2024SharingFeel}. Our findings extend previous results by showing that, beyond disclosing more over time, people's disclosures to robots are also more affective in nature and correspond to their emotion regulation practices. This aligns with affect labelling theory, suggesting that articulating emotions helps regulate them \cite{Torre2018PuttingRegulation,Lieberman2016PuttingWords}. 
Throughout the duration of their participation, participants enriched their disclosures with more descriptive language, incorporating a greater number of adjectives, potentially indicating deeper cognitive processing of their emotions and experiences. These findings may suggest that cognitive reappraisal with the robot appears to operate on two levels—changing interpretations while fostering cathartic expression—supporting the idea that verbalizing and reinterpreting emotions enhance regulation. Moreover, increased self-disclosure during reappraisal extends prior research on self-disclosure to robots \cite{Laban2024BuildingTime,laban_ced_2023, Laban2024SharingFeel}, further demonstrating its role in therapeutic improvement \cite{Farber2003PatientResearch,Farber2006Self-disclosurePsychotherapy}. Importantly, our results indicate that encouraging emotional expression during reappraisal does not overwhelm individuals but rather facilitates the process and improves outcomes. Future research should further validate this finding by evaluating whether the presence of the robot facilitating the intervention encouraged participants to engage in expression rather than suppression, or whether this effect was merely due to the task itself. Previous evidence from emotion regulation training \cite{Cohen2018FromTraining} and interpersonal emotion regulation research \cite{Marroquin2011InterpersonalDepression,Sahi2023PeerAdolescence,Zaki2013InterpersonalRegulation} indicates that incorporating a social element can significantly enhance cognitive reappraisal by engaging interpersonal processes that foster reflection and reinterpretation of emotional experiences. Accordingly, based on our quantitative and qualitative findings, we propose that the robot's social presence and interactivity may have positively impacted participants' emotion regulation performance in this experiment.

Interestingly, participants' expressed sentiment—verbal 
and non-verbal
—became gradually more negative within each session, even though their overall mood improved by the end. As conversations progressed, deeper reflection during the \textbf{NCQ} and \textbf{CR} disclosure items (see Section \ref{procedure}) appeared to prompt more negatively valenced speech and facial expressions, accompanied by heightened emotional arousal. 
This trend suggests that, rather than suppressing difficult emotions \cite{Butler2003TheSuppression}, participants engaged in reappraisal through genuine expression. This interpretation is supported by qualitative feedback, where participants noted that the robot helped them explore suppressed emotions and reconsider their perspectives. The observed pattern—initial emotional expression followed by cognitive restructuring—bridges a gap in emotion regulation theory, suggesting that effective reappraisal may involve allowing negative expressions to peak before relief is achieved \cite{Gross2003IndividualWell-Being,Gross2000ThePerspective}. These findings highlight a dynamic trajectory in robot-assisted emotion regulation, where encouraging expression, even of negative emotions, plays a crucial role in the regulation process. Our results support the premise that self-guided interventions \cite{Tong2024TheMeta-analysis,Edge2023TheMeta-analysis}, including technology-assisted interventions \cite{Sharma2024FacilitatingRestructuring} such as social robots \cite{Laban2024SocialWell-Being,RefWorks:404}, may help individuals navigate this dynamic mechanism while minimizing negative social consequences~\cite{Laban2024SharingFeel}. 

\subsection{Effects on Well-being}

The improvement in mood and increased positive sentiment after each session show that the 
intervention had a beneficial effect on participants' emotional states \textbf{(RQ4)}. While no cumulative effect on mood was observed over time, the consistent positive impact after each session underscores the intervention’s effectiveness in providing 
swift emotional support. For university students, who often face high levels of stress and emotional challenges due to academic pressures, personal development, and social expectations \cite{Park2020UnderstandingStudy.}, such swift and on-the-spot emotional uplift can be particularly valuable. Enabling students to access readily available supportive agents within their natural environments may help them better manage emotional difficulties as they arise, potentially preventing the escalation of stress into more serious mental health issues. These findings align with previous work showing that individuals experiencing negative emotions were more likely to engage in expressive interactions with a social robot, using the robot as an emotional outlet during distressing times \cite{Laban2023OpeningBehavior}. 

\subsection{Integration of Robotic Interventions in Familiar Settings}

By conducting the intervention in familiar environments within university halls and departments, the study aimed to simulate naturalistic interactions while maintaining a structured setting. This approach allowed students to engage with the robot in a setting that is part of their daily routines while ensuring experimental control. Accordingly, the robot provided accessible means for support with reappraisal and emotion regulation without requiring significant changes to their routines. Such a process may help students manage their emotions more effectively in real time, enhancing their ability to cope with stressors in their natural environments. The findings suggest that robot-assisted interventions can be a practical and scalable solution to support students' emotional well-being, offering emotional relief and fostering the development of long-term 
coping skills. Moreover, the study results underscore the broader potential of social robots, not only in emotion regulation but also in facilitating cognitive and behavioural change. 
From an affective computing perspective, our analysis of speech sentiment, facial expressions, and linguistic markers of self-reflection provides a framework for assessing users’ emotional engagement in real-world HRI. While our study applied this analysis post-hoc, it highlights how multi-modal emotion data can be leveraged to evaluate the effectiveness of interventions. Future affect-aware systems could build on this by exploring how changes in verbal and nonverbal 
cues correspond to shifts in emotion regulation over time. These findings add up to earlier work demonstrating how multimodal analysis of speech and sentiment in controlled HRI settings can reveal users' emotional engagement \cite{Laban2021}.

\section{Limitations and Future Research}

The study demonstrated promising outcomes, though several limitations should be considered. One limitation is the absence of a control group, which limits direct attribution of the effects solely to the intervention. However, in long-term HRI studies in real-world settings, control groups 
can be impractical due to environmental variability, inconsistent participant experiences, and logistical constraints \cite{Leite2013,10.1145/3412374}. To mitigate this, a within-subjects design was used, allowing participants to serve as their own controls \cite{Salkind2010Within-SubjectsDesign}, while LMEs accounted for both fixed and random effects, controlling for confounds over time \cite{Baayen2008Mixed-effectsItems,Bell2019FixedChoice}. This approach provides a robust framework for analysing the intervention’s impact in naturalistic settings while sustaining optimal power for explaining 
results \cite{Hesser2015ModelingInterventions}. Moreover, our sample size may be perceived as relatively small; however, this study was a long-term, mixed-methods investigation in everyday settings. The sample size was determined through an a priori calculation to ensure sufficient power \cite{RefWorks:410} and the data acquired and analysed aligns with previous HRI studies \cite{Spitale2024PastWell-being}. 
Finally, while the intervention promoted constructive emotion regulation strategies, it did not significantly reduce maladaptive ones. Future work could 
tailor interventions to minimise reliance on maladaptive strategies.

\section{Conclusions}

Our findings highlight the role of social robots not only in facilitating cognitive change but also in fostering richer emotional expression, which is central to effective emotion regulation. Notably, beyond the linear effects of increased expression over time, we also identified structured patterns of expressiveness aligned with the cognitive reappraisal process: participants expressed themselves more—and more positively in terms of sentiment and facial valence—within each session, with expression peaking during key reappraisal moments, particularly when the robot prompted them to reinterpret negative experiences using prior positive disclosures. This structured expressiveness underscores how robot-mediated reappraisal can guide users through a dynamic regulatory process, facilitating both emotional expression and cognitive change in a way that mirrors evidence-based therapeutic techniques. As technology advances, integrating social robots into daily environments could assist individuals in managing emotional challenges on a broader scale, offering an alternative approach to meeting the growing demand for social and emotional support in educational institutions and beyond.

\bibliographystyle{myIEEEtran}
\balance{\bibliography{ref_f}}
\begin{IEEEbiographynophoto}{Guy Laban} is a Postdoctoral Research Associate at the Department of Computer Science \& Technology of the University of Cambridge, and a member of the \href{https://cambridge-afar.github.io/}{Affective Intelligence and Robotics Laboratory (AFAR)}. Guy pursued his PhD studies in Neuroscience and Psychology as a Marie Skłodowska Curie Fellow at the School of Psychology and Neuroscience of the University of Glasgow. 
\end{IEEEbiographynophoto}

\begin{IEEEbiographynophoto}{Julie Wang} is a student at the Department of Psychology of the University of Cambridge, and a student intern at the \href{https://cambridge-afar.github.io/}{AFAR Lab}.
\end{IEEEbiographynophoto}

\begin{IEEEbiographynophoto}{Hatice Gunes} is a Full Professor of Affective Intelligence and Robotics (AFAR) in the Department of Computer Science and Technology, University of Cambridge, leading the \href{https://cambridge-afar.github.io/}{Cambridge AFAR Lab}. She is a former President of the Association for the Advancement of Affective Computing, a former Faculty Fellow of the Alan Turing Institute and is currently a Fellow of the EPSRC and Staff Fellow of Trinity Hall.
\end{IEEEbiographynophoto}


\end{document}